\documentclass[iicol,pdflatex,sn-mathphys-ay]{sn-jnl}

\usepackage{graphicx}%
\usepackage{multirow}%
\usepackage{amsmath,amssymb,amsfonts}%
\usepackage{amsthm}%
\usepackage{mathrsfs}%
\usepackage[title]{appendix}%
\usepackage{xcolor}%
\usepackage{textcomp}%
\usepackage{manyfoot}%
\usepackage{booktabs}%
\usepackage{algorithm}%
\usepackage{algorithmicx}%
\usepackage{algpseudocode}%
\usepackage{listings}%
\usepackage{subcaption}
\usepackage{algpseudocode}
\usepackage{float}

\theoremstyle{thmstyleone}%
\theoremstyle{thmstyletwo}%

\theoremstyle{thmstylethree}%

\begin{document}

\title[Article Title]{Volumetric density measurement in buoyant plumes using Tomographic Background Oriented Schlieren (TBOS)}


\author[1]{\fnm{Javed} \sur{Mohd}}\email{javedm@iitk.ac.in}

\author*[1,2]{\fnm{Debopam} \sur{Das}}\email{das@iitk.ac.in}

\affil[1]{\orgdiv{Department of Aerospace Engineering}, \orgname{Indian Institute of Technology Kanpur}, \orgaddress{\street{} \city{Kanpur}, \postcode{208016}, \state{Uttar Pradesh}, \country{India}}}

\affil[2]{\orgdiv{Department of Sustainable Energy Engineering}, \orgname{Indian Institute of Technology Kanpur}, \orgaddress{\street{} \city{Kanpur}, \postcode{208016}, \state{Uttar Pradesh}, \country{India}}}



\abstract{Buoyant plumes are encountered in both natural and engineered scenarios, ranging from volcanic ash clouds and wildfires to chimney smoke and industrial discharges into rivers and lakes. These plumes are driven by buoyancy forces arising from density differences between the plume and the surrounding fluid. However, measurements of the three-dimensional density field are scarce in the literature and absent for buoyant plumes. Such data can lead to a better understanding of pollutant dispersion, as well as the fundamental nature of concentration transport in three-dimensional and turbulent flows. We present three-dimensional density field measurements of buoyant plumes, particularly for lazy plumes, using an in-house-developed experimental rig and associated MATLAB-based processing codes. The density field is reconstructed using the Tomographic Background-Oriented Schlieren (TBOS) technique. The experimental rig consists of eight cameras facing random-dot backgrounds placed diametrically opposite, mounted circumferentially around the buoyant plume. Ray deflection (dot displacement) is calculated using a cross-correlation method. A finite-difference-based Poisson solver is employed to compute projected integrated density. Reconstruction is performed using the Simultaneous Algebraic Reconstruction Technique (SART). The resulting three-dimensional density field data is validated using similarity solutions and simulated data from the literature. We demonstrate that the 3D density field serves as valuable data for studying buoyant plume dynamics. For instance, in the case of lazy plumes, the measured density field clearly captures the puffing phenomenon. 
}




\maketitle
\section{Introduction}\label{sec:intro}
Many natural phenomena such as volcanic ash plumes \citep{andrews2009turbulent, bonadonna2015dynamics}, cloud thermals \citep{turner1969buoyant}, Earth's mantle plumes \citep{koppers2021mantle}, wildfire plumes \citep{finney2015role, kukkonen2014applicability, chung2023interaction}, hydrothermal vents at the ocean floor \citep{lupton1985entrainment}, glacial meltwater discharge into the sea \citep{hewitt2020subglacial}, and river discharging into the ocean \citep{horner2015mixing, rowland2009turbulent}  are driven by the buoyancy forces due to density difference between the ambient and plume fluids \citep{woods2010turbulent}. Similar buoyancy-driven flows are also encountered in several human-made scenarios, such as in heating and ventilation \citep{linden1990emptying, shrinivas2014transient,richardson2024flow}, industrial smoke or steam stack emissions \citep{slawson1967mean, raputa2023estimates}, waste-water discharge into the water bodies \citep{koh1975fluid, washburn1992mixing}, and plumes generated in deep-sea mining \citep{newland2024dynamics}. 

Accurate modelling of such phenomena is necessary to understand many of the catastrophic effects of these environmental events, related to natural disasters involving volcanoes, wildfires, and global warming. Industrial plumes must be discharged in a way that minimises their impact on nearby populations and ecosystems. However, as discussed below, the theoretical models for the prediction of such phenomena are still not very accurate, and a continuous effort is being put into improving them. Accurate measurement of the 3D density field could play a pivotal role in such efforts.

A theoretical model developed by Morton Taylor and Turner (MTT) \citep{morton1956turbulent} is widely used to predict the dynamics of such phenomena. Assuming `top-hat' radial profiles for vertical velocity and density, they employed conservation laws to derive governing equations that yielded similarity solutions. This initial model for buoyancy-driven flows was developed for idealised point sources and axisymmetric flow conditions. Several recent attempts have been made to extend this model to more realistic scenarios, including finite-area sources and non-Boussinesq plumes \citep{hunt2011classical, ciriello2020analytical}.

The theoretical model assumes that the entrainment velocity is proportional to the local vertical velocity along the plume axis to close the system of equations, leading to the use of a universal entrainment constant. However, several studies have shown that the assumption is inaccurate and reported a variable entrainment coefficient instead \citep{richardson2022entrainment, cenedese2024entrainment}. Therefore, the existing literature lacks a consensus on the value of the entrainment coefficient, which hinders the predictive capability of plume theories. Accurate estimation of the entrainment coefficient requires both velocity and density measurements. Apart from modelling the entrainment coefficient, several other models are being developed to describe the initial rise of plume \citep{turner1962starting, middleton1975asymptotic} and the dynamics of the plume head in both uniform and stratified surroundings \citep{richards2014radial, bhamidipati2017dynamics}. 

Often, these theoretical models need to be validated against the ground truth data that must be obtained from the measurements, as in the case of measuring entrainment. Therefore, to improve the predictive capabilities of the existing models or to develop new ones, one needs the accurate and detailed measurements of the plume parameters, including the full volumetric density field, for validation. 

The isothermal buoyant plumes, such as those in the current measurements, are governed by the conservation equations and the equation of state \citep{Bharadwaj_Das_2017} with unknown parameters of density, velocity, and pressure. Notably, if the 3D density measurements, as obtained by the developed technique in the current paper, are combined with simultaneous measurements of velocity, then the only remaining unknown, pressure, can be calculated. Thus, developing the 3D density measurement method is the first step toward achieving full parameter measurements in such buoyant plume problems. Moreover, simultaneous access to both density and velocity fields allows for the direct computation of the integral fluxes of volume ($Q = \iint_{A} (V \cdot n) \, dA$), momentum ($M= \iint_{A} V \rho  (V \cdot n) \, dA$), and buoyancy ($B = \iint_{A} g^ \prime (V.n) \, dA$) at any axial location without invoking geometric similarity assumptions- top hat or Gaussian profiles. The symbols $V$, $\rho$, $A$, $g^ \prime$ and $n$ denote the velocity field, density field, surface area through which flux is being calculated, reduced gravity and the unit normal to the surface, respectively. Consequently, the entrainment coefficient ($\alpha$) can be computed directly from the computed fluxes using the conservation of mass on a control volume \citep{bharadwaj2015near} at any axial location. This methodology provides a pathway to reconcile the wide discrepancies in entrainment coefficient values reported in the literature.

Most of the experimental measurements related to the buoyant plumes have been performed using liquid-in-liquid systems. Compared to liquid-based buoyant plume systems, gaseous buoyant plumes have received relatively less attention in experimental studies. A summary of such experimental measurements, including both liquid-based and gas-based systems, is provided in Table~\ref{tab:measSumry}.

\begin{table*}[h]
\caption{Summary of measurement techniques used in buoyant plume experiments.}\label{tab:measSumry}
\centering
\begin{tabular*}{\textwidth}{@{\extracolsep\fill}p{0.3\textwidth}p{0.7\textwidth}@{}} 
\toprule
Measurement Technique, Quantities & Author(s) \\ 
\midrule
Dye, Qualitative spread & \cite{batchelor1954heat, morton1956turbulent}, \cite{kitamura2011experiments}, \cite{kalita2024characterizing}, \cite{kaye2009experimental}, \cite{papanicolaou2020vertical} \\

Shadowgraphy/Schlieren, Density gradient & \cite{rogers2009natural} (dye), \cite{burridge2016fluxes}, \cite{webb2023turbulent} (thermocouple) \\

PIV , Velocity & \cite{sutherland2021plumes} \\

Scalar point measurements, Temperature, density & \cite{yuana1996experimental} (thermocouple), \cite{shabbir1994experiments} (hotwire), \cite{mi2001influence} (hotwire), \cite{rouse1952gravitational} \\

PIV + PLIF, Velocity and Species concentration & \cite{parker2020comparison}, \cite{paillat2014entrainment}, \cite{talluru2021turbulence}, \cite{milton2021entrainment}, \cite{wang2002second}, \cite{ai2006boussinesq}, \cite{o2005experimental} \\
\botrule
\end{tabular*}
\footnotetext{Source: This is an example of a table footnote.}
\end{table*}

As discussed above, the available measurement data on buoyant plumes remain predominantly point or two-dimensional, often limited to symmetric flow conditions in both liquid and gas-based buoyancy-driven flow systems. The current literature lacks three-dimensional density data for buoyant plumes. To address this gap, we employ Tomographic Background-Oriented Schlieren (TBOS) to measure the 3D density field in buoyant plumes. BOS measures the deflection of a ray due to light refraction within the measurement domain by taking two images of a background pattern: one with and one without the presence of a density gradient. This deflection corresponds to the path-integrated density gradient via the Gladstone-Dale relation between refractive index and density. The BOS measurement principle is detailed in Appendix \ref{appB:bosFunda}. To compute the absolute density of the measurement domain, path-integrated density gradients from multiple angular positions around it are tomographically combined using TBOS.

To date, 3D density measurements of buoyant plumes using TBOS have not been reported. Therefore, a summary of TBOS applications to other flow configurations is provided here. \cite{nicolas2016direct} reconstructed the density field in four convective flows: a candle flame, a hot jet generated by a heat gun, and two helicoidal flames obtained by rotation of a gas burner. \cite{nicolas20173d} also investigated various underexpanded jet flows issuing into quiescent air. \cite{hu2023background, hu2024reconstruction} measured temperature and density fields of single and dual laminar $CH_4$/air burner flames and helium jets in air, respectively. \cite{akamine2023formulation} obtained the 3D density of a candle flame placed near a wall using plane mirrors. \cite{unterberger2022evolutionary} measured the volumetric density of a Bunsen and co-flow burners using a premixed $CH_4$/air mixture with their evolutionary TBOS technique. \cite{liu2021volumetric} measured the density and temperature fields in a commercial Bunsen burner producing an asymmetric premixed butane-air flame. \cite{atcheson2008time} measured the density field in a gas burner using a 16-camera setup. \cite{amjad2020assessment, amjad2023three} investigated heated jets using a 15-camera setup, reporting the effect of different BOS parameters on the accuracy of reconstruction. \cite{grauer2018instantaneous} used a 22-camera experimental rig to reconstruct the three-dimensional refractive index field of an unsteady premixed natural gas/air flame from a Bunsen burner. The different variants of the TBOS used in the above-reported studies, and their relation to our TBOS variant, will be discussed later in this section.
Despite the application of TBOS to various flows\textemdash including combustion, heated jets and supersonic jets\textemdash 3D density field data for buoyant plumes remains, to the best of the authors' knowledge, unavailable in the literature. 

Before detailing the methodology of the present work, we briefly review existing approaches for reconstructing 3D density fields from BOS measurements. For a detailed review of the technique, readers are referred to \cite{raffel2015background, settles2017review, schmidt2025twenty}. The three-dimensional density in axisymmetric flows can be reconstructed by taking a single BOS recording from any viewing angle perpendicular to the direction of flow, followed by the reconstruction using an analytical technique, such as Filtered Back Projection (FBP) \cite{venkatakrishnan2004density} or Abel inversion \citep{champagnat2025abel}. In a second scenario, where the flow is stationary in time but asymmetric, a single camera BOS arrangement can be rotated around the measurement domain to obtain projections from multiple viewing angles, which can be used for 3D reconstructions. For all other generic cases: asymmetric and unsteady, multiple synchronised cameras must be used. For such flows available methods for the 3D density measurements, in the literature, can be broadly classified into three categories \citep{molnar2023estimating, schmidt2025twenty} (i) Indirect: classical three-step work-flows involving deflection sensing, Poisson integration, and tomographic reconstruction (the last two steps are not strictly implemented in the same chronological order), (ii) Direct: two-step methods that merge Poisson integration and reconstruction into a single inversion problem, and (iii) Unified: single-step approach that bypass intermediate displacement estimation altogether and combines all three steps in a single operator.

The classical three-step method, originally demonstrated by \cite{venkatakrishnan2004density}, and later also employed by \cite{atcheson2008time} and \cite{ota2011computed}, serves as a foundation for the present study. \cite{venkatakrishnan2004density} reconstructed the density field around a cone-cylinder arrangement in supersonic flow using a single camera view and Filtered Back Projection (FBP), leveraging the axisymmetry in the problem. They implemented a workflow pipeline comprising cross-correlation-based displacement estimation, 2D Poisson integration, and reconstruction via FBP. Our approach builds upon this workflow pipeline but adapts it to reconstruct unsteady and asymmetric buoyant plumes by using eight angular camera views and an iterative algebraic reconstruction technique (ART) variant, Simultaneous Algebraic Reconstruction Technique (SART) \citep{kak2001principles, grauer2020fast, davis2021tomographic, jia2024tomographic}. This plane-by-plane reconstruction strategy reduces computational load by limiting the size of the system matrix and solution vector. Furthermore, it makes use of extensively validated components readily available in the literature, such as `PIVlab' for computing cross-correlations, Poisson solver, and SART algorithm. However, other implementations of the three-step method, such as those by \cite{atcheson2008time} and \cite{ota2011computed}, first perform voxel-based 3D reconstruction instead of plane-by-plane reconstruction, and then perform Poisson integration on the full 3D reconstructed gradient field to obtain the density or refractive index.

For the methods of indirect and direct type, the deflection of the background pattern has to be estimated. This is usually achieved by taking a brightness consistency constraint between the two images, with and without the presence of density gradients. The methods, such as window-based cross-correlation, Optical Flow (OF) \citep{lucas1981iterative, HORN1981185}, and Fourier demodulation that uses a periodic background pattern \citep{wildeman2018real, shimazaki2022background} are employed to compute the deflection field \citep{schmidt2025twenty}. OF methods have gained popularity in modern BOS processing pipelines due to their higher spatial resolution, typically producing one displacement vector per pixel. However, they require tuning of regularisation parameters, which is non-trivial in the absence of ground truth data, and are computationally intensive, often an order of magnitude costlier than cross-correlation approaches \citep{schmidt2025twenty}. Nevertheless, OF remains the deflection-sensing method of choice in many recent studies. Moreover, \cite{schwarz2023practical} reported that different OF algorithms were unable to automatically achieve the same resolution and accuracy as a cross-correlation-based technique with an interrogation window size of 12x12. 

Among direct approaches, \cite{nicolas2016direct, nicolas20173d} proposed a method that combines Poisson integration and tomographic inversion steps into a unified framework, using OF-based displacement fields as input. While this approach is now commercially available, direct comparative studies with the classical three-step method, to the best of the authors' knowledge, are not available. The reduction in error between the three-step and two-step methods was inferred theoretically, without support from empirical data. The method’s reliance on the \cite{lucas1981iterative} paradigm of computing OF, where a window is selected as a user input under the assumption that the pixels within it have the same velocity, makes it equivalent to a cross-correlation-based approach. The two-step direct approach increases memory demand due to solving the full 3D domain simultaneously, which enlarges both the system matrix and the solution vector. In contrast, the current plane-by-plane reconstruction limits the solution vector to unknowns only on the plane under consideration, thereby reducing memory requirements. However, both the two-step and current approach solve the inverse problem using iterative methods-conjugate gradient and SART, respectively, which avoid the active storage of the full system matrix in memory; instead, the system-matrix solution-vector product is computed on the fly. This is achieved using a function that calculates the $i$-th row of the system matrix on demand to multiply it with the solution vector.

\cite{grauer2020fast} introduced Unified BOST (UBOST), a single-step reconstruction framework that directly uses image intensities, with and without flow, as input, eliminating the explicit displacement estimation step. The technique combines all three steps regarding deflection computation, inversion, and Poisson integration into a single operator. This is achieved by substituting the unknown deflections in the brightness consistency constraint with the deflection defined by BOS projections (forward projection  model) through the phase object. The unknown density of the phased object is then recovered by inverting the system derived from this substitution. They further empirically compare the accuracy of UBOST with the two-step approach (BOST), and report that UBOST produces reconstructions with a higher correlation coefficient relative to the ray-traced exact solution. They also showed that UBOST converged in 62.5\% of the time taken by two-step BOST, at most. 

However, the lack of empirical data comparing the three-step reconstruction approach with the two-step approach limits the estimation of the accuracy gain between them. In contrast, the relative accuracy gain from the two-step to the one-step approach has been empirically reported as a cross-correlation value of approximately 0.95 to 0.96 \citep{grauer2020fast}. If a similar accuracy gain is assumed when transitioning from the three-step to the two-step approach, the reconstruction quality of the three-step method should be considered adequate. Nevertheless, an empirical comparison is necessary to quantify the accuracy improvement between the three-step and two-step approaches.

Parallel to these developments, several data-driven methods based on Convolutional Neural Networks (CNNs) and Physics-Informed Neural Networks (PINNs) have been proposed \citep{cai2021flow, mucignat2023lightweight, molnar2023estimating, rudenko2025complete, rudenko2025reconstruction, teh2025indoor}. These approaches aim to further improve reconstruction fidelity. However, data-driven approaches also require training using ground truth data, which is typically generated through high-fidelity Computational Fluid Dynamics (CFD) simulations, though these simulations may themselves be subject to computational errors. Moreover, the range of real-world measurement conditions can be vast, and capturing all such scenarios by a trained model will require huge computational resources, as noted in \cite{schmidt2025twenty}. Furthermore, they point out that if high-fidelity training datasets can be reliably generated with high accuracy using CFD, the need for TBOS-based measurements using such data-driven, neural network-based models may not be justifiable. BOS-derived data are increasingly being used for flow data assimilation \citep{cai2021flow, rohlfs2024assimilating} in computational simulations.

A quantitative comparison of our methodology with the alternatives (such as the direct two-step and unified single-step methods) in terms of reconstruction accuracy is beyond the scope of this work. The main contribution of the present study is to extend the classical three-step BOS reconstruction framework of \cite{venkatakrishnan2004density} to asymmetric, unsteady flows\textemdash specifically, to obtain 3D density fields in buoyant plumes. 

The present work provides the first 3D density measurements of buoyant plumes with varying source strengths. Furthermore, the volumetric density field is utilised to visualise the dynamics of puffing buoyant plumes. This work builds upon our previous work \cite{mohd2022development}, where we presented the results of the proof-of-concept stage. Here, we extend the analysis by refining the technique, including the cubic spline sinogram interpolation, considering weights generated in the reconstruction circle only, and performing computation on a very dense grid using a Hamming window. We developed an in-house buoyant plume measurement rig based on the work of \cite{Bharadwaj_Das_2017, bharadwaj2019puffing}, along with the MATLAB\textsuperscript{\textregistered}-based code to process BOS images and tomographically reconstruct the 3D density field.

The remainder of this paper is organised as follows: Sect.~\ref{sec:methodology} describes the methodology, including experimental setup, Poisson integration, forward tomographic modelling, and SART reconstruction. In Sect.~\ref{sec:results}, we validate the measurements using similarity solutions and available simulation data in the literature, followed by the visualisation of the puffing plume dynamics. Finally, conclusions and future directions are presented in Sect.~\ref{sec:conc}. The details of the BOS measurement principle (Appendix~\ref{appB:bosFunda}), camera alignment and coordinate system validation (Appendix~\ref{append:camAlignErrors}), and uncertainty quantification (Appendix~\ref{appC:errorEstimation}) arising from background pattern selection, blur circle, exposure time, lens distortions, and parallel beam forward modelling are discussed. The Appendix \ref{app:quasiShadowEffects} discusses the quasi-shadowgraph effects due to background illumination passing through the phase object. The computation of ray deflection using cross-correlation is detailed in Appendix~\ref{appB:crossCorr}. The effect of sinogram interpolation and post-processing steps is detailed in Appendices~\ref{append:D}.

\section{Methodology}\label{sec:methodology}
This section describes the experimental rig, including the buoyant plume generation (Sect.~\ref{subsec:setupPlumGen}) and TBOS imaging (Sect.~\ref{subsec:meth_bosCamArrangement}) setups. Using this experimental rig, we obtain the raw images of the background with and without the plume and process them using an in-house software suite to obtain the volumetric density field. The workflow is described in Sect.~\ref{subSec:dataWorkflow}. 

\subsection{Buoyant plume generation and characterization}\label{subsec:setupPlumGen}
The buoyant plume experimental rig consists of two subsystems: the apparatus to generate the desired buoyant plume and the TBOS imaging setup. 

\begin{figure}[h]
\centering
\includegraphics[width=0.49\textwidth]{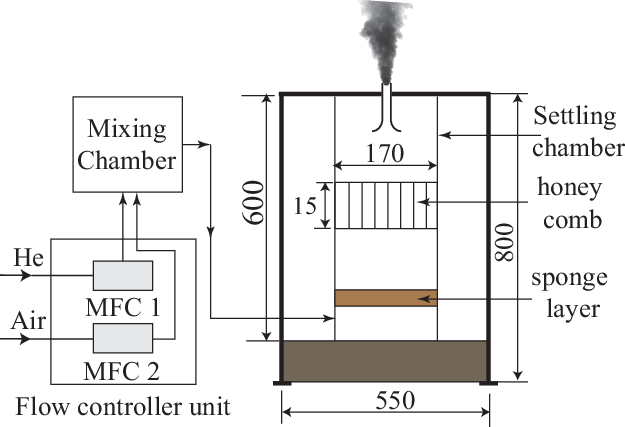}
\caption{Schematic of experimental setup for generating buoyant plumes. All dimensions are in mm.}
\label{fig:plumeSetup} 
\end{figure}

The experimental setup for the generation of the buoyant plume is shown schematically in Fig.~\ref{fig:plumeSetup}. Two mass flow controllers, MFC 1 and MFC 2 (Alicat Scientific, MFC-100 SLPM, MFC-1000 SLPM), are used to precisely intake the dry helium (He) and air gases, which are being supplied from the compressed gas cylinders. Inside a specially designed mixing chamber fitted with silencers \citep{Bharadwaj_Das_2017}, two gases are mixed, then fed to the vertically mounted settling chamber. The settling chamber has a sponge layer and a honeycomb structure that helps to damp out any remaining flow disturbances and to straighten the flow. The He-Air mixture finally passes through a pipe nozzle and discharges into the ambient air. The nozzle has an exit diameter of $D = 35$ mm and a length of 314 mm, featuring a bell-mouth inlet (diameter $2D$) to ensure smooth flow entry. The exit of the nozzle is mounted at the centre of a heavy steel bench platform. For further details of plume generation setup, refer to \citep{Bharadwaj_Das_2017}; however, the current setup does not include the seeding mechanism, as this is not required. The cuboid acrylic enclosure has also been removed for the ease of installing cameras and background patterns around the setup as described in Fig.~\ref{fig:bosSetUp}. Experiments are performed in a quiet room. 

Given the flow rates $Q_{\text{He}}$ (SLPM) and $Q_{\text{Air}}$ (SLPM), taking the constant temperature and pressure during mixing, the equivalent source volume flow rate $Q_0 = Q_{\text{He}} + Q_{\text{Air}}$, and the equivalent mass flow rate $m_0 = \rho_{\text{He}} Q_{\text{He}} + \rho_{\text{Air}} Q_{\text{Air}}$ are calculated. The reference properties of pure He and air gases taken for the analysis are presented in Table~\ref{tab:gasRefParams}. Further, the equivalent density of the mixture gas is given as $\rho_0 = m_0/Q_0$. The equivalent dynamic viscosity, $\mu_0$, of the mixture is calculated using Wilke's formula \citep{wilke1950viscosity}. The exit velocity at source, $V_0 =Q_0/\pi b_0^2$, where $b_0=D/2$ is the source radius. The buoyant plumes are characterised using the parameters listed in Table~\ref{tab:cases}.

\begin{table}[h]
\caption{Properties of helium (He) and air used in the analysis}\label{tab:gasRefParams}%
\begin{tabular}{@{}lll@{}}
\toprule
Property & He  & Air \\
\midrule
density ($\rho$), kg/m$^3$   & 0.1635   & 1.184   \\
dynamic viscosity ($\mu$), Pa$\cdot$s     & 1.984e-5  & 1.849e-5   \\
molecular weight ($m_w$), g/mol       & 4.0026   & 28.966   \\
Gladstone-Dale constant ($G$)\footnotemark[1], m$^3$/kg    & 1.960e-04 & 2.239e-4 \\
\botrule
\end{tabular}
\footnotetext{Experiments were conducted at 298 K temperature and 1 atm pressure.}
\footnotetext[1]{Data from \cite{merzkirch2012flow}.}
\end{table}

\begin{table*}[h]
\caption{Characterization of the buoyant plumes at source (denoted by subscript `0').}
\label{tab:cases}
\centering
\begin{tabular}{@{}p{0.06\textwidth} p{0.025\textwidth} p{0.025\textwidth} p{0.025\textwidth} p{0.025\textwidth} p{0.03\textwidth} cccccc@{}}
\toprule
Release & \multicolumn{2}{c}{$\dot{Q}$ (SLPM)} & $Y_{\text{He}}$ & $Y_{\text{Air}}$ & $\rho_0$ & $V_0 $ & $S_0$ & $Re_0$ & $Fr_0$ & $Ri_0$ & $\Gamma_0 \, $ \\
\cmidrule(lr){2-3} 
     & He & Air & & & $\frac{kg}{ m^{3}}$ & $\frac{m}{s}$ & & & & &  \\
\midrule
I & 20 & 10 & 0.22 & 0.78 & 0.50 & 0.52 & 0.43 & 449 & 0.89 & 1.72 & 3.12 \\
II & 20 & 15 & 0.16 & 0.84 & 0.60 & 0.61 & 0.51 & 633 & 1.03 & 0.91 & 1.97 \\
III & 30 & 10 & 0.29 & 0.71 & 0.42 & 0.69 & 0.35 & 494 & 1.18 & 1.31 & 1.98 \\
\botrule
\end{tabular}
\end{table*}

The parameters given in Table~\ref{tab:cases} are the mass fractions $Y$, density ratio $S=\rho_0 / \rho_{\infty}$, Reynolds number $Re=\rho_0 V_0 D/ \mu_{0}$, Froude number $Fr=V_0/ \sqrt{gD}$, and Richardson number $Ri=(\rho_{\infty} - \rho_0) g D / \rho_{0} V_{0} ^2$. The symbols $\rho_0$, $\rho_{\infty}$, $V_0$, and $D$ denote the density of source, the density of ambient air, velocity at source, and diameter at source, respectively. The subscript `0' denotes the value of a parameter at the source. The flux balance parameter (or the plume function), denoted as $\Gamma_0$ and defined for the finite area source plume \citep{morton1973scale, hunt2011classical}, is calculated as $(5 B_0 Q_0^2)/(8 \alpha \sqrt{\pi} M_0^{5/2})$, where $B_0$ is the source buoyancy flux, $ Q_0 $ is the volume flux, $ M_0$ is the momentum flux, and $ \alpha $ = 0.091 (for Gaussian profiles, \cite{ciriello2020analytical}) is the entrainment coefficient. Based on the source plume function ($\Gamma_0$), plume can be characterised \citep{hunt2011classical} to be forced ($0<\Gamma_0<1$), pure ($\Gamma_0 =1$), and lazy ($1< \Gamma_0$). 

A relation between density ($\rho$) and the refractive index ($n$) of the He-Air mixture can be derived \citep{qin2002effect,Ghazwani2016, wanstall2020implications, hu2024reconstruction} using the Gladstone–Dale relation along with the ideal gas law for mixtures, given as:   
\begin{align}
 \label{eq:gladstone-combined}
n - 1 &= \frac{P}{R_u T} \left( G_{\text{He}} \, m_{\text{wHe}} \, \chi_{\text{He}} + G_{\text{Air}} \, m_{\text{wAir}} \,\chi_{\text{Air}} \right) \\
\rho &=  \frac{P m_w}{R_u T} \\
m_w &= m_{\text{wHe}} \, \chi_{\text{He}} + m_{\text{wAir}} \, \chi_{\text{Air}} \\
1 &= \chi _ {\text{He}} + \chi_{\text{Air}}  ;
\end{align}

where $P$, $R_u$, and $T$ denote the pressure, universal gas constant, and temperature, respectively. $G$, $m_w$, and $\chi$ denote the Gladstone–Dale constant, molecular weight, and molar fraction. The subscripts indicate whether the quantity pertains to helium (He) or air. The values of $G$ and $m_w$ are taken from Table~\ref{tab:gasRefParams}. The final relation between the mixture density gradient and the refractive index gradient can be obtained by eliminating $\chi$, resulting in: 

\begin{equation}
\nabla \rho = \left(\frac{ G_{\text{He}} m_{\text{wHe}} - G_{\text{Air}} m_{\text{wAir}} }{m_{\text{wHe}} -  m_{\text{wAir}}} \right)^{-1} \nabla n.
\end{equation}

Using the respective values of He and Air parameters, the equivalent Gladstone-Dale constant for the mixture of He-Air is obtained as 2.2837 $\times 10^{-4}$ m$^3$/kg.

\subsection{Identification of puffing and non-puffing cases}\label{subsec:meth_idPuffNonPuff}
Using the neutral curves for buoyant plume stability from the global instability analysis in \cite{Bharadwaj_Das_2017, kkbPhDThesis}, we classify the three release cases in the present study as either laminar or puffing. Puffing buoyant plumes are characterised by the periodic formation of toroidal vortices near the source, which convect downstream. These structures are also responsible for flame flicker and pinch-off in candle flames~\citep{maxworthy1999flickering,xia2018vortex}.

\begin{figure}[h]
\centering
\includegraphics[width=0.49\textwidth]{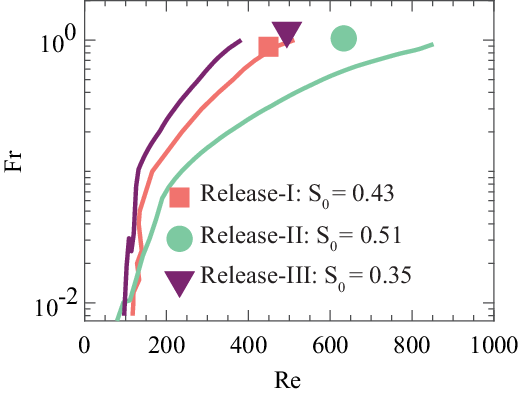}
\caption{Classifying three plume releases into laminar or puffing regimes using the non-dimensional parametric space of Re-Fr. Solid curves taken from \citep{Bharadwaj_Das_2017} are the neutral stability curves of the primary mode obtained from the global instability analysis and separate the laminar (stable) and puffing (unstable) regions.}
\label{fig:neutralCurve} 
\end{figure}

In Fig.~\ref{fig:neutralCurve}, the three solid curves correspond to the density ratios ($S$) of the release cases listed in Table~\ref{tab:cases}. The Reynolds number ($Re$) and Froude number ($Fr$) for each case are plotted as filled symbols. A case is considered stable (laminar) if the point lies to the left of its corresponding neutral curve, and unstable (puffing) if it lies to the right. Furthermore, the farther a point lies from the neutral curve, the stronger the degree of stability or instability it represents. Based on this classification, one case (Release-II) is identified as laminar-non puffing, while the other two (Releases-I and III) are unstable and exhibit puffing behaviour, as seen in Fig.~\ref{fig:neutralCurve}. Release-I lies close to the neutral curve, suggesting it is near the transition between laminar and puffing regimes. In contrast, Release-III lies farther to the right of its stability curve, indicating a higher degree of instability and, hence, stronger puffing behaviour compared to Release-I. These stability characteristics will be referenced in the results section to validate our volumetric density field measurements.

\subsection{BOS camera arrangement}\label{subsec:meth_bosCamArrangement}
\begin{figure*}[ht]
\centering
\includegraphics[width=0.95\textwidth]{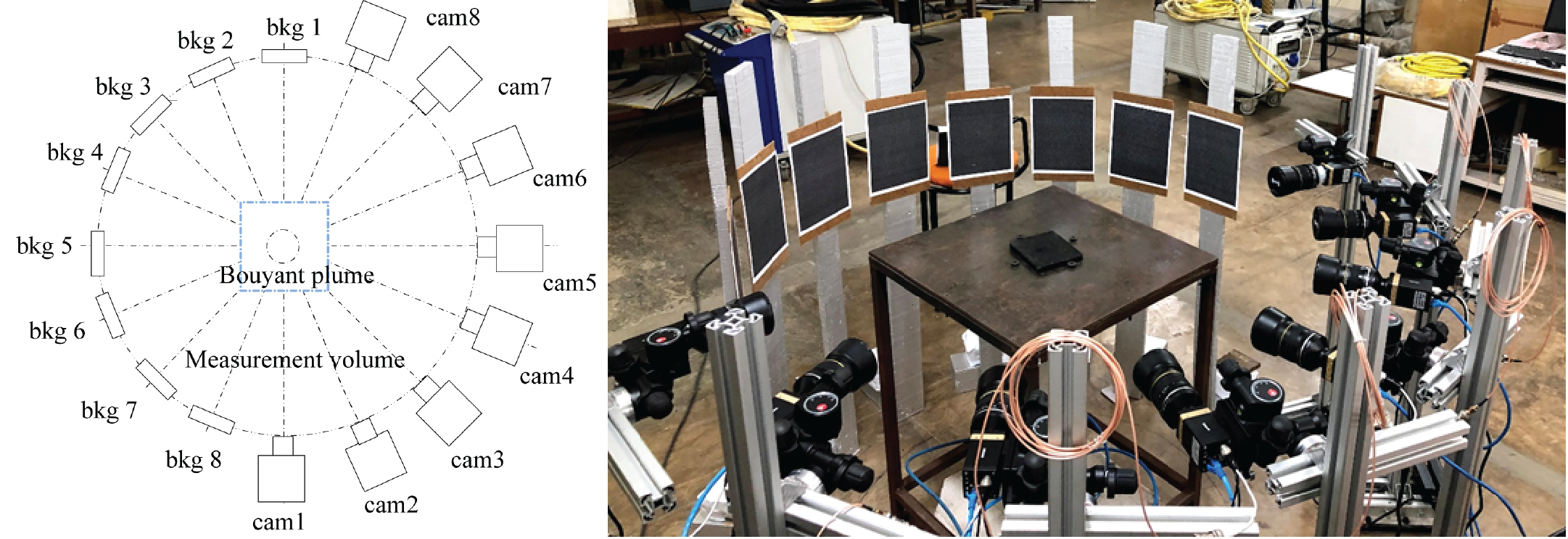}
\caption{8-camera laboratory setup for Tomographic BOS, centred around the plume: (a) schematic and (b) actual experimental setup. The symbols `cam' and `bkg' denote the camera and background, respectively.}
\label{fig:bosSetUp} 
\end{figure*}

Figure \ref{fig:bosSetUp} depicts the schematic of the camera arrangement for tomographic BOS measurements of the buoyant plume. Eight IMPERX GEV-B2320M-TC000 cameras (2352 x 1768 px) with 85 mm lenses, each facing a background pattern, are placed on a circle around the buoyant plume in such a way that the line-of-sight of the camera and background passes through the centre of the measurement volume. The radius of the circle and angular spacing are 600 mm and $22.5^{\circ}$, respectively. The details of the camera alignment procedure are presented in Appendix \ref{append:camAlignErrors}. The camera Field Of View (FOV) and the tomographic reconstruction square size calculations are discussed in Appendix \ref{appC:errorEstimation}, which also details the important aspects of the camera setup that affect BOS imaging accuracy, such as the centre of confusion, exposure time, and lens distortions.

The background patterns are illuminated using a single, 1000 W halogen lamp with a color temperature range around 3400 K. The characterization of the background patterns is provided in Appendix \ref{appC:errorEstimation}. It is mounted on the camera side of the rig, positioned sufficiently high above the measurement domain to avoid quasi-shadowgraph effects \citep{schmidt2025twenty} caused by the illumination light passing through the phase object. The details for the assessment of quasi-shadowgraph effects are presented in Appendix~\ref{app:quasiShadowEffects}. All eight cameras are synchronized using a hardware trigger signal obtained from a BNC Model 575 digital/delay pulse generator at 10 fps for Release-I and II and 21 fps for Release-III, with an exposure time of 100~$\mu s$ for acquiring the images simultaneously. The acquisition and processing were performed using a Dell Precision Tower 7810 equipped with an Intel(R) Xeon(R) CPU E5-2650 v3, 2.30GHz (10 cores), 64.0 GB of RAM, and an NVIDIA Quadro K5000 (4 GB) graphics card, running on a 64-bit Windows 10 Pro operating system. The cameras, operating on the GigE Vision protocol, were connected using CAT6 Ethernet cables to two dedicated Network Interface Cards (NICs), with each NIC handling four cameras. The camera control and image acquisition were performed using Bobcat GEV Software and IMPERX ToolKit.  

\subsection{Data Processing work-flow}
\label{subSec:dataWorkflow}
\begin{figure*}[ht]
\centering
\includegraphics[width=0.95\textwidth]{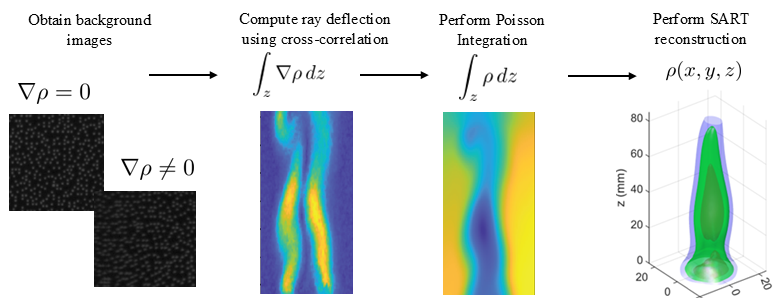}
\caption{Data processing chain of the developed TBOS system }
\label{fig:workflow}
\end{figure*}

The raw images obtained from the BOS experimental rig, both in the presence and absence of plumes (i.e., density gradients), are shown in the left part of Fig.~\ref{fig:workflow}. These images, captured from each of the eight cameras, are subjected to further processing. In step one, the ray deflection is computed using a cross-correlation-based approach, resulting in a 2D line-integrated density gradient, typically referred to as projection data in tomographic literature. The obtained data are further subjected to Poisson integration in step two, yielding a line-integrated density. Finally, in step three, tomographic reconstruction is performed using the SART. For the SART reconstruction, one row of data from each of the 2D line-integrated density projections, obtained from each of the eight camera views, is used to form a sinogram. A sinogram is a two-dimensional representation of the projection data for a particular plane. The sinogram is then used as input to the SART algorithm to reconstruct the density field in the target plane. All the density planes are subsequently stacked to form a three-dimensional density field. The data processing workflow is illustrated in Fig.~\ref{fig:workflow}, and the details of each of the three steps are presented in the following sections.  

\subsubsection{Ray deflection estimation using cross-correlation}\label{subsec:xCorr}
\begin{figure*}[ht]
\centering
\includegraphics[width=0.95\textwidth]{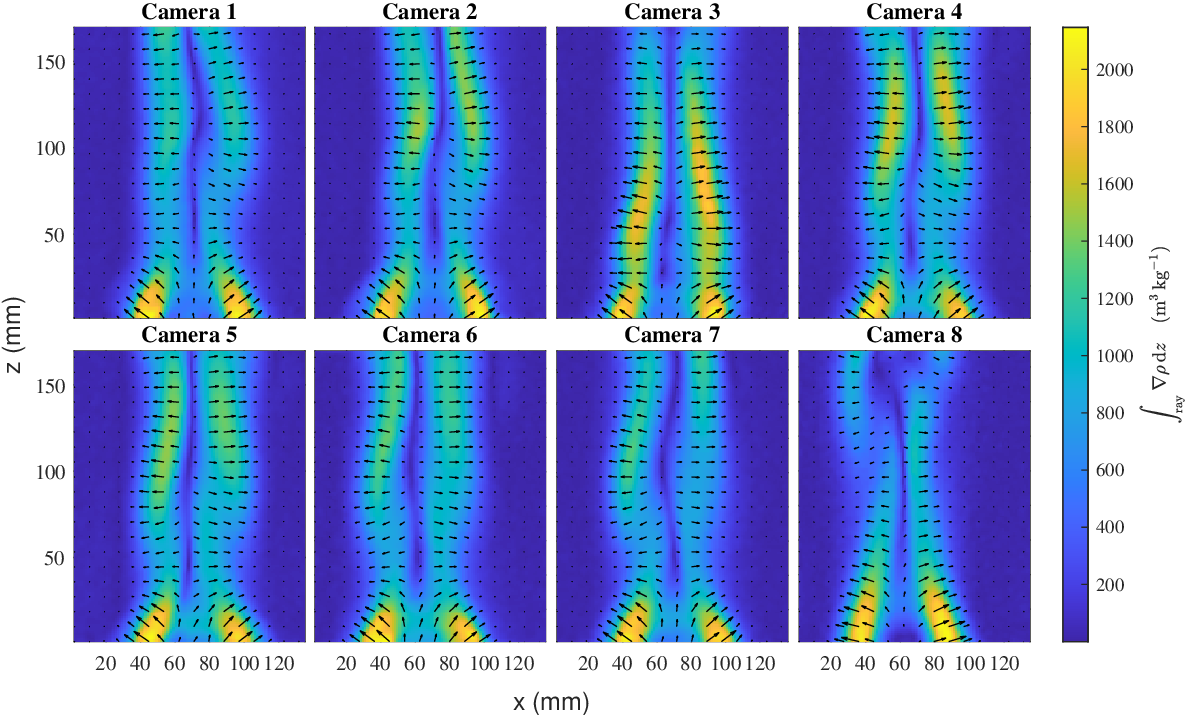}
\caption{Projected density gradients from eight cameras at background plane (scale: 0.0773 mm/px).}
\label{fig:denGrad}
\end{figure*}

The ray deflection magnitudes obtained from processing the BOS images using the cross-correlation algorithm, as detailed in Appendix~\ref{appB:crossCorr}, for Release-I (Frame-3) from all eight cameras are shown in Fig.~\ref{fig:denGrad}. 

\subsubsection{Poisson solver to obtain line integrated density field}\label{subsub:PoisnSolver}
A 2D Poisson integration is performed on the deflection field obtained from each camera using an FDM-based, in-house developed Poisson solver. The projected density gradients ($\int \nabla \rho \, dz$), as given by Eq.~(\ref{eq:bosFinal}) for the y-direction, are obtained by applying cross-correlation to the background images with and without the plume. In this step of the reconstruction workflow, the projected density ($\int\rho \, dz$) is obtained from projected density gradients by formulating and solving a Poisson problem.

Consider the Poisson equation in 3D for a scalar-valued function $ \psi(x, y, z) $, given by:

\begin{equation}
    \nabla^2 \psi = \frac{\partial^2 \psi}{\partial x^2} + \frac{\partial^2 \psi}{\partial y^2} + \frac{\partial^2 \psi}{\partial z^2} = \phi(x, y, z),
\label{eq:poisson}
\end{equation}
where $ \phi(x, y, z) $ represents the source term. This equation is discretized using the finite difference method (FDM) on a 3D Cartesian grid with uniform spacing $ h_1 $, $ h_2 $, and $ h_3 $ in the $ x $-, $ y $-, and $ z $-directions, respectively. For an interior grid point $ (i, j, k) $, the second-order central differences in each spatial direction (7-point stencil with 3 points in each direction) yield the following discretized equation:

\begin{align}
\frac{\psi_{i+1,j,k} - 2\psi_{i,j,k} + \psi_{i-1,j,k}}{h_1^2} + \nonumber \\
\frac{\psi_{i,j+1,k} - 2\psi_{i,j,k} + \psi_{i,j-1,k}}{h_2^2} + \nonumber \\
\frac{\psi_{i,j,k+1} - 2\psi_{i,j,k} + \psi_{i,j,k-1}}{h_3^2} = \phi_{i,j,k}.
\label{eq:discrete_poisson}
\end{align}

The discretised equation, when applied to each grid point (node), results in one equation. All the equations obtained from each node can be rearranged to form a linear system (Eq.~(\ref{eq:linear_system})). In the linear system, unknown $\psi$ values are rearranged to form the vector $\underline{\mathbf{\psi}}$ and the matrix $\mathbf{B}$ includes the coefficients of those unknown values. Therefore, Eq.~(\ref{eq:discrete_poisson}) when applied to each internal node results in one row of the linear system,
\begin{equation}
    \mathbf{B} \underline{\mathbf{\psi}} = \underline{\mathbf{\phi}},
    \label{eq:linear_system}
\end{equation}
where $ \mathbf{B} $ is a sparse matrix and $ \underline{\mathbf{\phi}} $ is the vector of source terms. 

To formulate the Poisson problem with projected density gradients, the source function vector $\underline{\mathbf{\phi}}$, defined in Eq.~(\ref{eq:poisson}), is calculated  as:
\begin{equation}
\underline{\mathbf{\phi}}  = \frac{\partial \Delta_x}{\partial x} + \frac{\partial \Delta_y}{\partial y}+\frac{\partial \Delta_z}{\partial z},
\end{equation}
where $\Delta_x$ and $\Delta_y$ are calculated using cross- correlation in PIVlab and the gradient in the optical direction is taken as $\Delta_z = 0$.

Neumann boundary conditions are implemented using a second-order accurate central difference stencil, assuming fictitious (ghost) points. The value at the ghost node ($\psi_{i+1}$) is defined in terms of the interior node ($\psi_{i-1}$) and the measured ray deflection ($\Delta_x$) as $\psi_{i+1} = \psi_{i-1} + 2h_1 \Delta_x$. In the context of this paragraph, the index $i$ denotes the boundary point. This expression is substituted into the discretised Poisson equation at the boundary node to eliminate the ghost node, resulting in a modified source term at the boundary: $\phi_i = \frac{h_1^2}{2} (\frac{\partial \Delta_x}{\partial x})_i \pm h_1 \Delta_{x,i}$, where the sign of the second term is determined by the orientation of the boundary normal. Boundary faces with normals in the $y$ and $z$ directions are treated analogously. The boundary conditions are incorporated into the system by modifying the matrix $\mathbf{B}$ and the right-hand side vector $\underline{\mathbf{\phi}}$ in Eq.~(\ref{eq:linear_system}). Subsequently, a Dirichlet boundary condition, $\psi = \psi_{0}$, is imposed later, after the reconstruction step (see Sect.~\ref{subSec:3D-Recon}), to correct the undetermined integration constant in the Poisson solution arising from the exclusive use of Neumann boundary conditions.

We use MATLAB function `[$\mathbf{L},\mathbf{U},\mathbf{P},\mathbf{Q},\mathbf{D}$] = \texttt{lu}($\mathbf{B}$)' to perform a sparse LU decomposition, which factorizes the matrix $ \mathbf{B}$ as:
\begin{equation}
    \mathbf{P} \left( \mathbf{D} \backslash \mathbf{B} \right) \mathbf{Q} = \mathbf{L} \mathbf{U},
\end{equation}

where $\mathbf{L}$ and $\mathbf{U}$ are lower and upper triangular matrices, respectively, $\mathbf{P}$ and $\mathbf{Q}$ are permutation matrices, and $\mathbf{D}$ is a diagonal scaling matrix that improves the stability and sparsity of the factorization. The solution process consists of two steps. First, forward substitution is used to solve for an intermediate vector $\underline{\mathbf{y}} $ in the system $L \underline{\mathbf{y}} = \mathbf{P} \underline{\mathbf{\phi}}$. Then, backward substitution is applied to solve for the final solution $\underline{\mathbf{\psi}}$ in the system $U \underline{\mathbf{\psi}} = \underline{\mathbf{y}}$. By utilizing row scaling via $\mathbf{D}$, the factorization remains sparse and stable, allowing efficient computation for large-scale systems.

\begin{figure*}[h]
\centering
\includegraphics[width=0.95\textwidth]{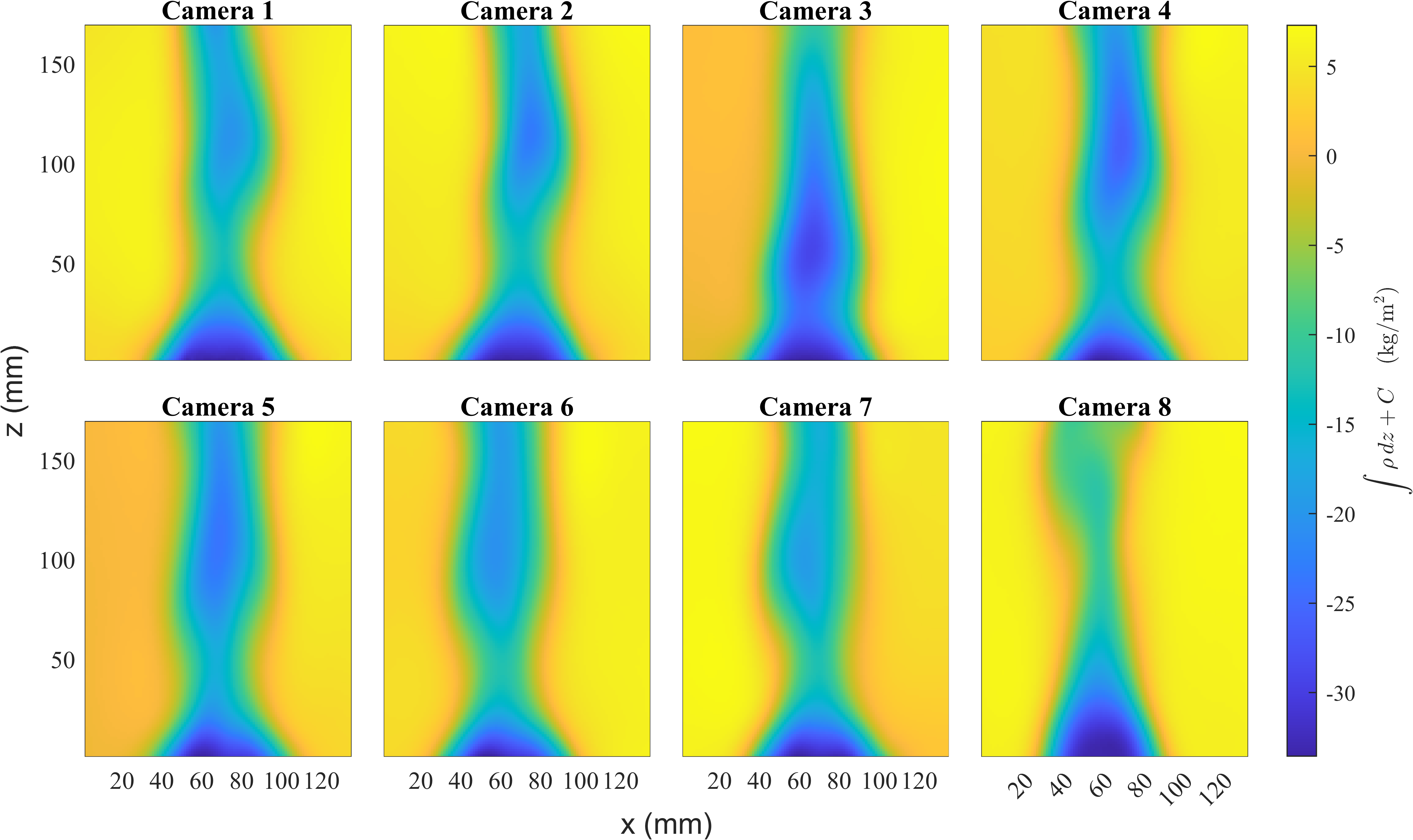}
\caption{Projeced Poisson-integrated density fields from eight cameras at background plane (scale: 0.0773 mm/px).}
\label{fig:lineIntDen}
\end{figure*} 

Although the solver is natively 3D, a 2D Poisson integration is employed by replicating the 2D projection data across three adjacent grid layers in the $z$-direction. In this configuration, the central plane functions as the solution layer, while the two flanking planes serve as ghost layers. This creates a computational volume where the solution is invariant along $z$ ($\psi_{i,j,k-1} = \psi_{i,j,k} = \psi_{i,j,k+1}$), thereby forcing the discrete second derivative $\partial^2/\partial z^2$ to vanish. Consequently, the 3D discretised Laplacian reduces exactly to its 2D counterpart. 

The solution of the Eq.~(\ref{eq:linear_system}) yields the projected density fields, $\int \rho dz+C$, where $C$ denotes the remaining integration constant, for each of the 8 cameras (Fig.~\ref{fig:lineIntDen}). These results from the Poisson solver are further used to reconstruct  density in each transverse plane, $\rho(x, y)$, using a tomographic reconstruction algorithm, as described in the following sections. The reconstructed planes are then stacked to form the full 3D density field, $\rho(x, y, z)$. 

\subsubsection{Sinogram Interpolation}
\label{sub:sinInt}
\begin{figure}[h]
\centering
\includegraphics[width=0.48\textwidth]{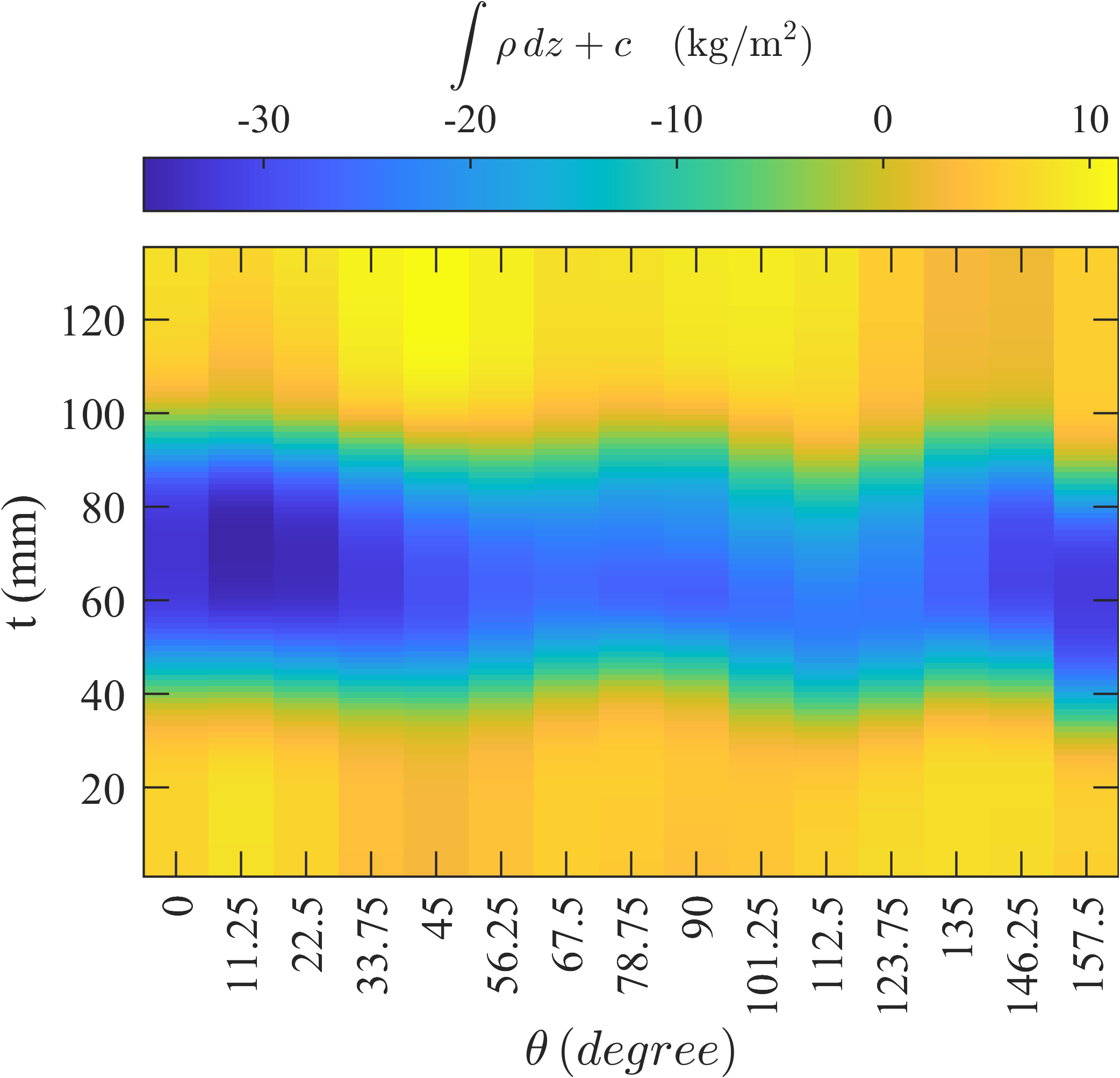}
\caption{Interpolated sinogram showing synthesized data between original measurements at eight angles. Odd-numbered ticks represent experimental data at 22.5$^\circ$ intervals, while even-numbered ticks indicate interpolated values derived from adjacent true measurements.}
\label{fig:Sinogrm}
\end{figure}

Sinogram inpainting, or interpolation, has been shown to improve the reconstruction quality \citep{brooks1978new} by eliminating the streak artefacts \citep{kalke2014sinogram} in sparse angle tomography. Various interpolation methods exist, ranging from geometric approaches \citep{bertram2004directional, constantino2008sinogram, li2011strategy} to data-driven techniques, including machine learning and deep learning-based methods \citep{li2019sinogram, yao2024no, bellens2024machine}. 
To improve the accuracy \citep{li2011strategy} of the reconstruction from sparse angle tomographic measurements, the sinogram from eight projection angles is interpolated using cubic spline interpolation, following the approach by \cite{enjilela2019cubic}. The 2D discrete sinogram data was interpolated using MATLAB's `interp2' function, which incorporates cubic spline interpolation. After the interpolation, the number of angular views artificially increases from 8 to 15. A representative interpolated sinogram is shown in Fig.~\ref{fig:Sinogrm}. The interpolation introduces synthesized data between experimentally acquired views, effectively presenting the tomographic reconstruction algorithm with 15 projection angles. In this work, reconstruction using 15 angles with the SART algorithm was observed to produce fewer artefacts than using the original eight projection angles (Appendix \ref{append:D}). However, it should be noted that the synthesized data introduced through interpolation assumes a smoothly varying density field between adjacent angular measurements. This assumption is justified, as the plumes, though asymmetric, are laminar in nature and are expected to exhibit smoothly varying features across the azimuthal angular spacing of 22.5$^\circ$.

\subsubsection{Tomographic Reconstruction using SART} \label{subSec:3D-Recon}

The forward projection model for the tomographic reconstruction process is first defined, followed by the methods used to solve the inverse problem. A parallel beam tomographic model is employed due to small divergence errors, as quantified in Appendix \ref{appSec:FovAndRayDivErr}.

\begin{figure}[h]
\centering
\includegraphics[width=0.45\textwidth]{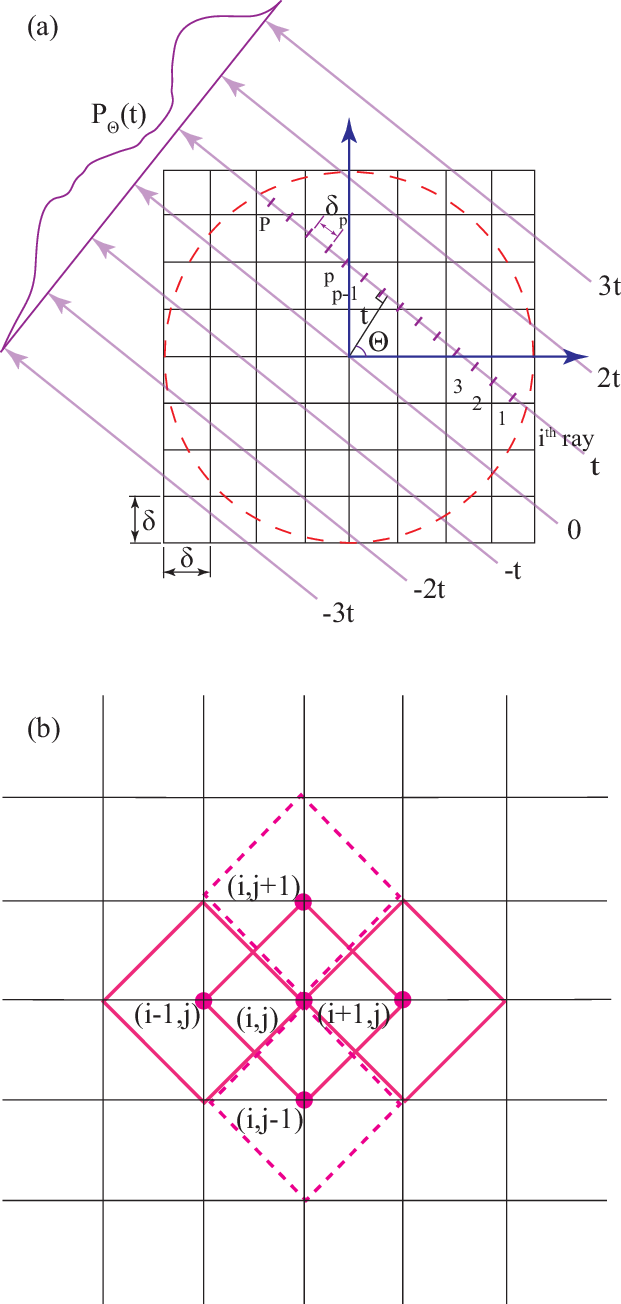}
\caption{Forward Projection Modelling (a) Radon Transform schematic and (b) The Pyramid shape bilinear basis function of unit height with support till the  adjacent 4 nodes }
\label{fig:forwardProj} 
\end{figure}

A projection is defined as the line integral of the function f(x,y) along a ray located at a perpendicular distance t from the origin and oriented at an angle $\theta$ counter-clockwise from the $x$-axis (Fig.~\ref{fig:forwardProj}a). Each ray is uniquely defined by parameters- t and $\theta$. For all points (x,y) on the ray, t = xcos($\theta$) + ycos(90-$\theta)$. The projection along the ray ($\theta$, t) is defined as $p_{\theta}(t) = \int_{\text{line}(\theta, t)} f(x,y)\, ds$, which can be expressed in the Radon transform form as $\int \int_{\mathbb{R}^2} f(x, y) \, \delta\left( t - (x \cos \theta + y \sin \theta) \right) \, dx \, dy $. Denoting the Radon transform operator corresponding to ray $(\theta,t)$, indexed by i, as $\mathcal{R}_i$, the projection becomes $=\mathcal{R}_i f(x,y)$.

Assuming that the function can be approximated by N basis functions $\phi(x,y)$, such that $f(x,y)\approx \tilde{f}(x,y)=\sum  _{j=1} ^N f_j \phi_j(x,y)$, where $f_j$ are the coefficients associated with each basis function. Therefore, the Radon transform, in terms of discretized function f(x,y) can be written as $p_i =\mathcal{R}_i \tilde{f}(x,y)$, substituting discrete approximation of $\tilde{f}(x,y)$ yields $=\mathcal{R}_i \sum  _{j=1} ^N f_j \phi_j(x,y)$ $\rightarrow$ $=\sum  _{j=1} ^N \{ \mathcal{R}_i   \phi_j(x,y) \}  f_j $
 $=\sum  _{j=1} ^N a_{ij}  f_j$ where, $a_{ij}$ denotes the scaler obtained by applying operator $\mathcal{R}_i$ on $j^{th}$ basis function $\phi_j(x,y)$. 

The line integral for Radon transform can also be calculated by sum of discrete values of function $\tilde{f}(s_{ip})$ at points $1<p<P$ at  regular intervals $\delta_p$ (Fig.~\ref{fig:forwardProj}a) on the ray inside the domain - $p_i = \sum_{p=1}^{P}  \tilde{f}(s_{ip}) \, \delta_p$ where, the values $\tilde{f}(s_{ip})$ are calculated using the bilinear interpolation of the basis function surrounding the $p^{th}$ point on $i^{th}$ ray. For ease of forming linear system of equation, it is written as a bilinear contribution from all the shape functions- $\tilde{f}(s_{ip})  = \sum _{j=1}^{N} w_{ijp} f_j$, where $w_{ijp}$ is the weight contribution from j$^{\textit{th}}$ basis function $\phi_j$ to the p$^{\textit{th}}$ point on i$^{\textit{th}}$ ray (Fig.~\ref{fig:forwardProj}a). However, only the four basis functions surrounding the point contribute, since the support of each pyramid-shaped basis function is limited to its four neighboring grid nodes (Fig.~\ref{fig:forwardProj}b).

Now, substituting $\tilde{f}(s_{ip})$ in $p_i = \sum_{p=1}^{P}  \tilde{f}(s_{ip}) \, \delta_p$ results: $p_i = \sum_{p=1}^{P}   \sum _{j=1}^{N} w_{ijp} f_j \, \delta_p$ $\rightarrow$
$=  \sum _{j=1}^{N} \left\{ \sum_{p=1}^{P}    w_{ijp}  \, \delta_p \right\} f_j$. 
Comparing this to the $p_i=\sum  _{j=1} ^N a_{ij}  f_j$, the coefficients $a_{ij}$ can be obtained as 

\begin{equation}
a_{ij}=\sum_{p=1}^{P}    w_{ijp}  \, \delta_p
\label{eq:aijDef}
\end{equation}

Therefore, the final linear system representing the forward projection modelling is given by
\begin{equation}
\mathbf{A} \underline{\mathbf{f}} = \underline{\mathbf{p}}
\label{eq:linSystForward}
\end{equation}
where $\mathbf{A} \in \mathbb{R}^{M \times N}$ is the matrix with coefficient $a_{ij}$ given by Eq.~(\ref{eq:aijDef}), $ \underline{\mathbf{f}} \in \mathbb{R}^{N \times 1}$ is the vector of unknowns or the solution vector, and $\underline{\mathbf{p}} \in \mathbb{R}^{M \times 1}$ is the projection vector obtained by vectorizing the sinogram of the target plane. Whereas, the sinogram is formed by extracting the projection data rows, corresponding to the target plane, from the solution of Eq.~(\ref{eq:linear_system}). This inverse problem (Eq.~(\ref{eq:linSystForward})) will be solved using Kaczmarz methods of solving algebraic equations \citep{kaczmarz1937, kak2001principles}.

During the implementation, the coefficients $a_{ij}$ of the matrix $\mathbf{A}$ are computed using $\delta_p = \delta / 8$, where $\delta = \texttt{domSize} / \texttt{nPixRecon} = 0.44~\mathrm{mm}$. Here, \texttt{domSize} ($48.51~\mathrm{mm}$) and \texttt{nPixRecon} ($110$) refer to the reconstruction square side length and the number of pixels along the side, respectively. The value of \texttt{domSize} is computed from the maximum field of view (FOV) width ($W_{md} = 68.6~\mathrm{mm}$) as $W_{md} / \sqrt{2}$, as detailed in Appendix~\ref{appC:errorEstimation}, Sect.~\ref{appC:camFOV}, and Fig.~\ref{fig:camFOV}.

The analytical tomographic reconstruction technique FBP, based on the Fourier slice theorem, though computationally faster, suffers from poor accuracy when there are limited angular projections. On the other hand, the ARTs, which are iterative, provide better reconstruction accuracy with limited projections and noisy data \citep{kak2001principles}. In this study, the SART, a variant of ART, is employed to develop an in-house code for reconstructing the instantaneous density field from BOS measurements. Unlike ART, where the solution $\underline{\mathbf{f}}$ is updated after the consideration of each ray, SART averages the corrections for all the rays and uses the averaged value to update the solution. This completes one iteration, and the process continues until a suitable convergence criterion is met. The iteration step in SART can be mathematically represented as follows:
 
\begin{equation}
 \label{eq:SART_hann}
\underline{\mathbf{f}}^{(k+1)} =\underline{\mathbf{f}}^{(k)} + \frac{1}{M} \sum_{i=1}^{M} \left( \frac{p_i - \underline{\mathbf{a}}_i \cdot\underline{\mathbf{f}}^{(k)}}{\underline{\mathbf{a}}_i \cdot \underline{\mathbf{a}}_i} \right) \underline{\mathbf{t}}_i,
 \end{equation} 
 
where the vector of unknowns at $k^{\textit{th}}$ iteration is $\underline{\mathbf{f}}^{(k)} = [f_1, f_2, \dots, f_N]^T$, $\underline{\mathbf{a}}_i = [a_{i1}, a_{i2}, \dots, a_{iN}]$ represents the $i^{\text{th}}$ row of the matrix $ \mathbf{A} $ with elements $a_{ij}$, and $\underline{\mathbf{t}}_i = [t_{i1}, t_{i2}, \dots, t_{iN}]$ is the $i^{\text{th}}$ row of the matrix with elements $t_{ij}$. $N$ is the total number of unknowns or grid points, and $M$ is the total number of projection rays. The initial guess for the solution vector $\underline{\mathbf{f}}^{(0)}$is set to $\underline{\mathbf{0}}$.

The elements $t_{ij}$ are calculated as:
 \begin{equation}
t_{ij}=\sum_{p=1}^{P} h_{ip} \,    w_{ijp}  \, \delta_p,
\label{eq:tijDef}
\end{equation}
where $h_{ip}$, for $1 < p < P$, is the Hanning window applied along the ray $i$ over $P$ discrete points.  The Hanning window, similar to a Hamming window, which was found to give superior results \citep{andersen1984simultaneous, kak2001principles}, has been employed.

In the SART, the iterative reconstruction process is terminated when the relative percentage change in the solution vector norm falls below a predefined threshold, $\epsilon = 4\%$. The stopping criterion is defined as $\|\underline{\mathbf{f}}^{(k)} -\underline{\mathbf{f}}^{(k-1)}\|_2 \, / \, \|\underline{\mathbf{f}}^{(k)}\|_2  \times 100 < \epsilon$, where $\underline{\mathbf{f}}^{(k)} $ and $\underline{\mathbf{f}}^{(k-1)} $ represent the current and previous solution vectors, respectively, and $\|\cdot\|_2$ denotes the Euclidean norm (i.e., $\|\underline{\mathbf{f}}\|_2 = \sqrt{\sum_{i=1}^n f_i^2}$). The threshold $\epsilon = 4\%$ was empirically chosen based on multiple trials of reconstruction quality to ensure physical consistency, including centre-line Gaussian profiles, necking of the lazy plume, and the convection of the structures. This criterion monitors the stability of the solution vector, which represents the reconstructed image, ensuring convergence while preventing over-iteration that could amplify noise \citep{andersen1984simultaneous, warnett2025don}. Though a more formal, quantitative method for determining the optimal stopping point may further improve reconstruction accuracy. In the future, probably L-curve \citep{hansen1992analysis} or any other suitable method \citep{hansen2010discrete} could be used. Though typically used for explicitly regularized inversions, the L-curve method can also be used for unregulated iterative inversions, where the number of iterations plays the role of regularization parameter \citep{hansen2010discrete}.

After the reconstruction, the axis grids- x, y, and z are scaled down to the measurement domain's scaling factor (mm/pixel), which is half the scaling factor at the background pattern location. Furthermore, the density field is post-processed to correct the undetermined integration constant due to the exclusive use of Neumann boundary conditions in the Poisson integration step as described in Sect.~\ref{subsub:PoisnSolver}. A Dirichlet boundary condition with source density $\rho_0$ for each case at the first reconstruction plane at the nozzle exit is used to correct for the integration constant. Next, values outside the circular region (shown by dashed red circle in Fig.~\ref{fig:forwardProj}a) ($x^2 + y^2 \leq (L/2)^2$, $L$ being the square domain size) were set to $\rho_{\text{air}}$ to remove the noise in the corners of the reconstruction domain. Also, values where $\rho > \rho_{\text{air}}$ are set to $\rho_{\text{air}}$ to enforce the physical constraint; such values were found only near the circumference of the circle defined above. Finally, a 3D smoothing filter is applied using MATLAB’s `smooth3' function with the `box' method, which performs a moving average over each cubical window of size 5x5x5 grids, reducing the high frequency noise. A detailed discussion on the effect of each post-processing step defined here is presented in Appendix~\ref{append:D} and Fig.~\ref{fig:sngInterpAndPostProcessing}.

\begin{figure}[h]
\centering
\includegraphics[width=0.5\textwidth]{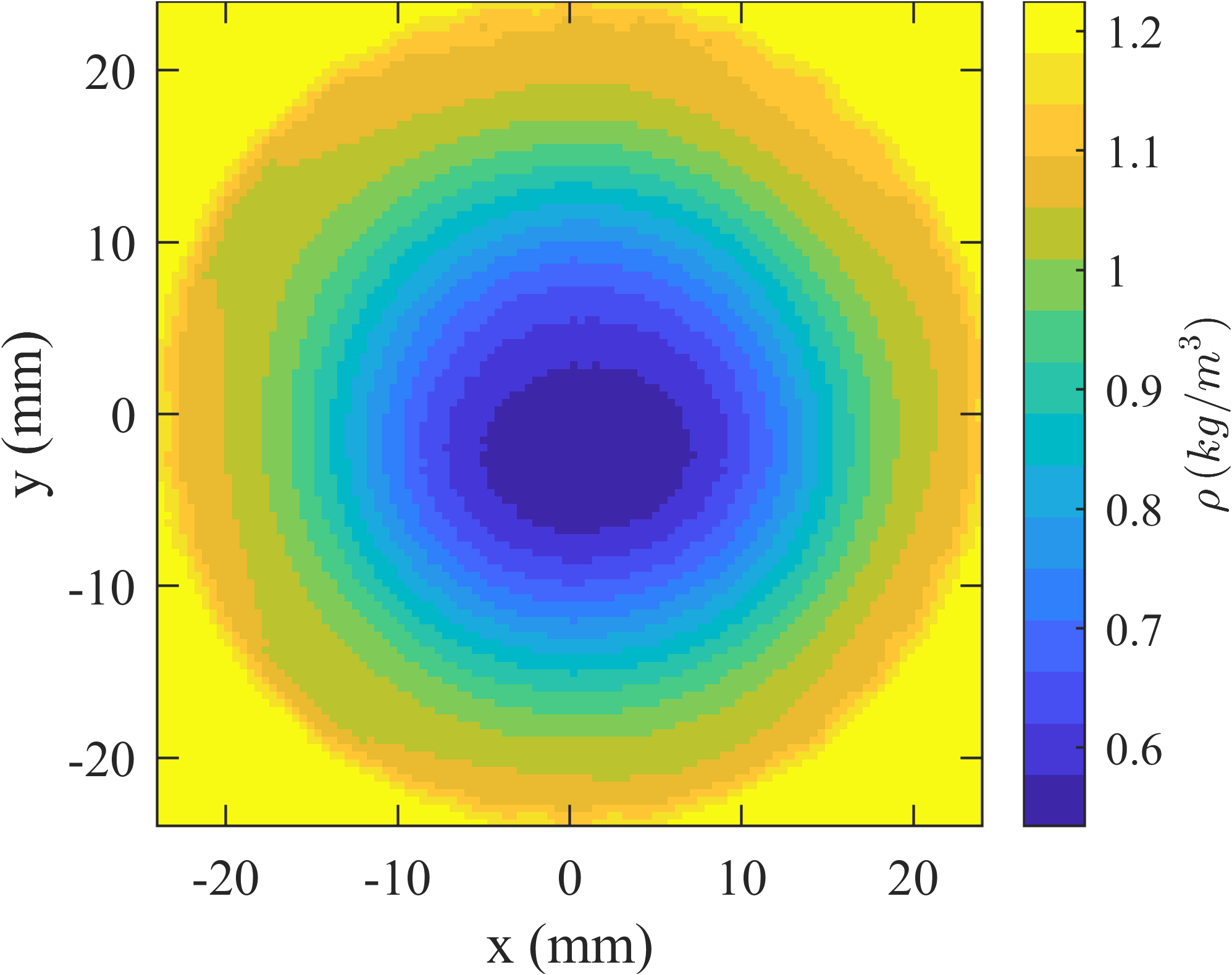}
\caption{A representative instantaneous reconstructed density plane extracted at an arbitrarily selected time instance (snap-3) from the reconstructed density field of Release-I at a height of 6.17 mm (plane-10) from the source (scale: 0.0773/2 mm/px).}
\label{fig:denPlaneDemo}
\end{figure}

A representative post-processed instantaneous reconstructed density plane for Release-III at an arbitrary selected time instance is shown in Fig.~\ref{fig:denPlaneDemo}.
\section{Results}\label{sec:results}
The reconstructed three-dimensional density field is validated by comparing key features with expected plume behaviour, such as Gaussian transverse profiles and the similarity solutions for reduced gravity and plume radius along the downstream z-direction. Following the validation, we present visualizations of the puffing phenomenon in buoyant plumes based on the reconstructed 3D density fields.

\subsection{Validation using plume theory}\label{subsec:resValid}
\begin{figure*}[ht]
\centering
\includegraphics[width=0.7\textwidth]{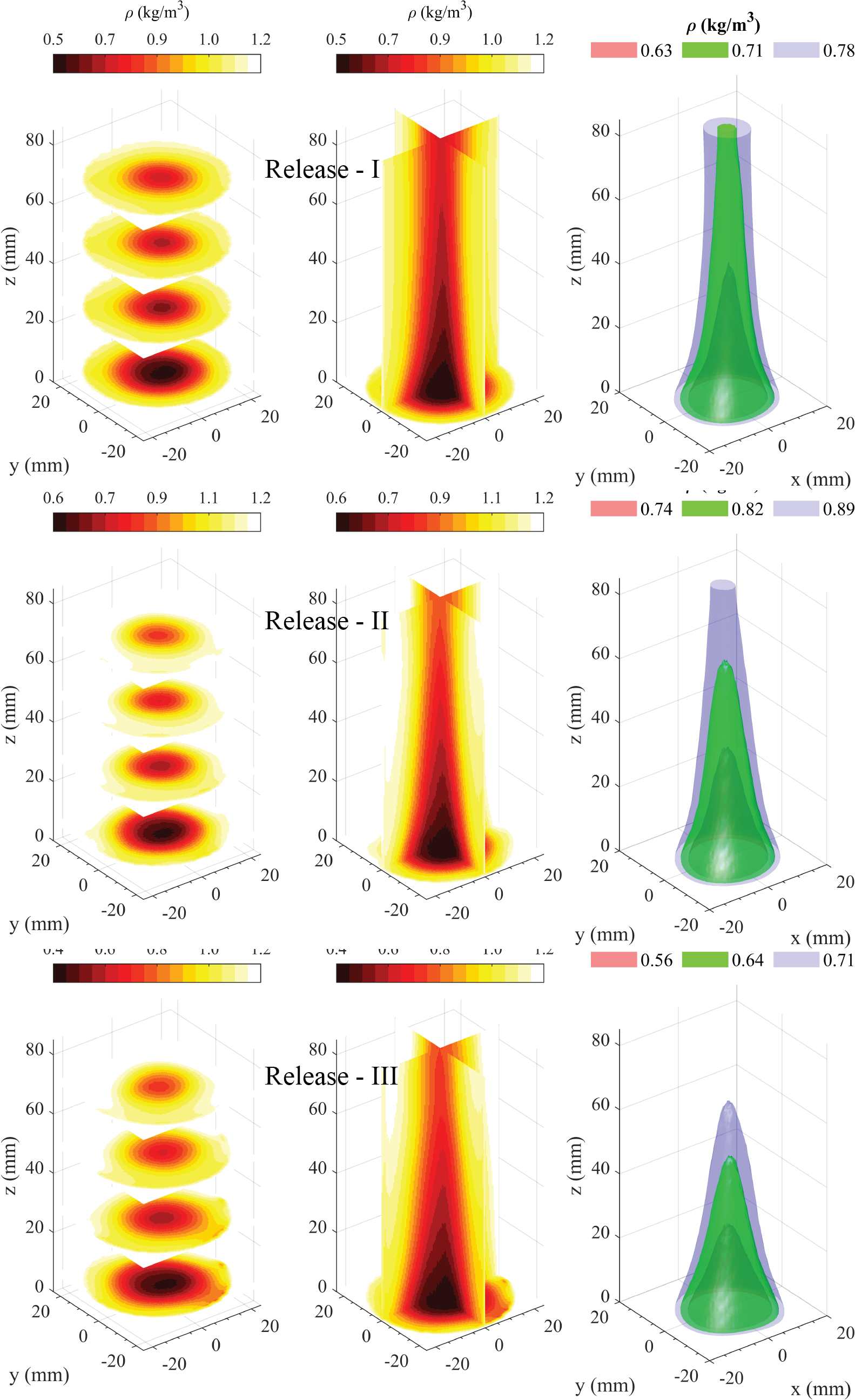}
\caption{Volume visualization of time-averaged density for all three releases: I (top row), II (middle row), and III (bottom row) at the measurement domain (scale: 0.0773/2 mm/px). The right column shows iso-surface visualizations, the middle column features longitudinal density planes that highlight internal variations within the iso-surfaces, and the left column presents transverse density slices to capture radial variations. }
\label{fig:validatn} 
\end{figure*} 

The instantaneous 3D density fields obtained for all three releases (Table~\ref{tab:cases}) are time averaged and shown in Fig.~\ref{fig:validatn}. A total of 33, 29, and 55 images were used for averaging in Release-I, II, and III, respectively. The instantaneous reconstructions (see supplementary videos for each release case) span multiple puffing cycles, resulting in a well-converged time average. A few time instances had corrupted frames due to acquisition errors, leading to the exclusion of corresponding camera images and minor temporal gaps in the BOS sequence. These omissions cause slight visual discontinuities in the videos but do not affect the analysis or interpretation of the flow features.

The left column in each case shows individual planes as reconstructed using our TBOS processing code. The middle column shows the density field after stacking all the planes together and then taking slices through the axis of the plume. The right column shows the iso-surfaces at selected values. Three iso-surfaces are drawn at increasing density levels, with the innermost and outermost iso-surfaces corresponding to the lowest and highest values, respectively, as indicated in the colourbars. As can be seen, all three cases show the lowest density fluid near the source region (represented by red color). The density of the fluid increases as we move radially outward and in the downstream direction due to entrainment of heavier ambient fluid, air, into the plume. This is evident by the presence of blue, the largest density, iso-surface in the outermost region and in the downstream region of the plume.

\subsubsection{Transverse Density Profile Characterization: Top-Hat vs. Gaussian}
 To validate the density measurements, $\rho (r,z)$, the time-averaged field is non-dimensionalized as reduced gravity, $g^\prime$. Next, the $g^\prime$ profiles extracted at different downstream axial locations, above the source, are examined for self-similarity. For a circular plume emanating from a finite area source of diameter D, the flow variable  $g^\prime (r,z)$ follows a Gaussian distribution \citep{ezzamel2015dynamical} as: 
\begin{align}
g'(r,z) &= g'_c(z) \exp \left( \frac{-r^2}{b_{g'}^2} \right), 
\label{eq:gaussProfiles}
\end{align}
where 
\begin{equation}
g' = \left( \frac{\rho_\infty - \rho}{\rho_\infty} \right) g.
\label{eq:gDash}
\end{equation}

$g'_c(z)$ and $b_g'$ are the centreline value and the coefficient of radial scaling for reduced gravity, respectively. 

\begin{table*}[h]
\caption{Gaussian fit parameters}
\label{tab:gausFitParams}
\centering
\begin{tabular}{@{}c |cccc| cccc| cccc@{}}
\toprule
Release & \multicolumn{4}{c}{ I ($ S_0 =0.43, \, Ri_0=1.72$)} & \multicolumn{4}{c}{ II ($ S_0 =0.51, \, Ri_0=0.91$)} & \multicolumn{4}{c}{ III ($ S_0 =0.35, \, Ri_0=1.31$)} \\ 
\midrule
$z/D$ & 0.18 & 0.79 & 1.41 & 2.03 & 0.18 & 0.79 & 1.41 & 2.03 & 0.18 & 0.79 & 1.41 & 2.03 \\ 
$g'_c$ & 5.59 & 4.55 & 4.12 & 3.82 & 4.76 & 3.75 & 3.15 & 2.57 & 6.23 & 4.91 & 4.27 & 3.60 \\ 
$b_{g'}$ (mm) & 14.86 & 12.13 & 11.61 & 11.84 & 12.60&9.62 & 8.46 & 8.03 &  15.47 & 12.67 & 11.16 & 9.67 \\ 
\botrule
\end{tabular}
\end{table*}

\begin{figure*}[ht]
\centering
\includegraphics[width=0.9\textwidth]{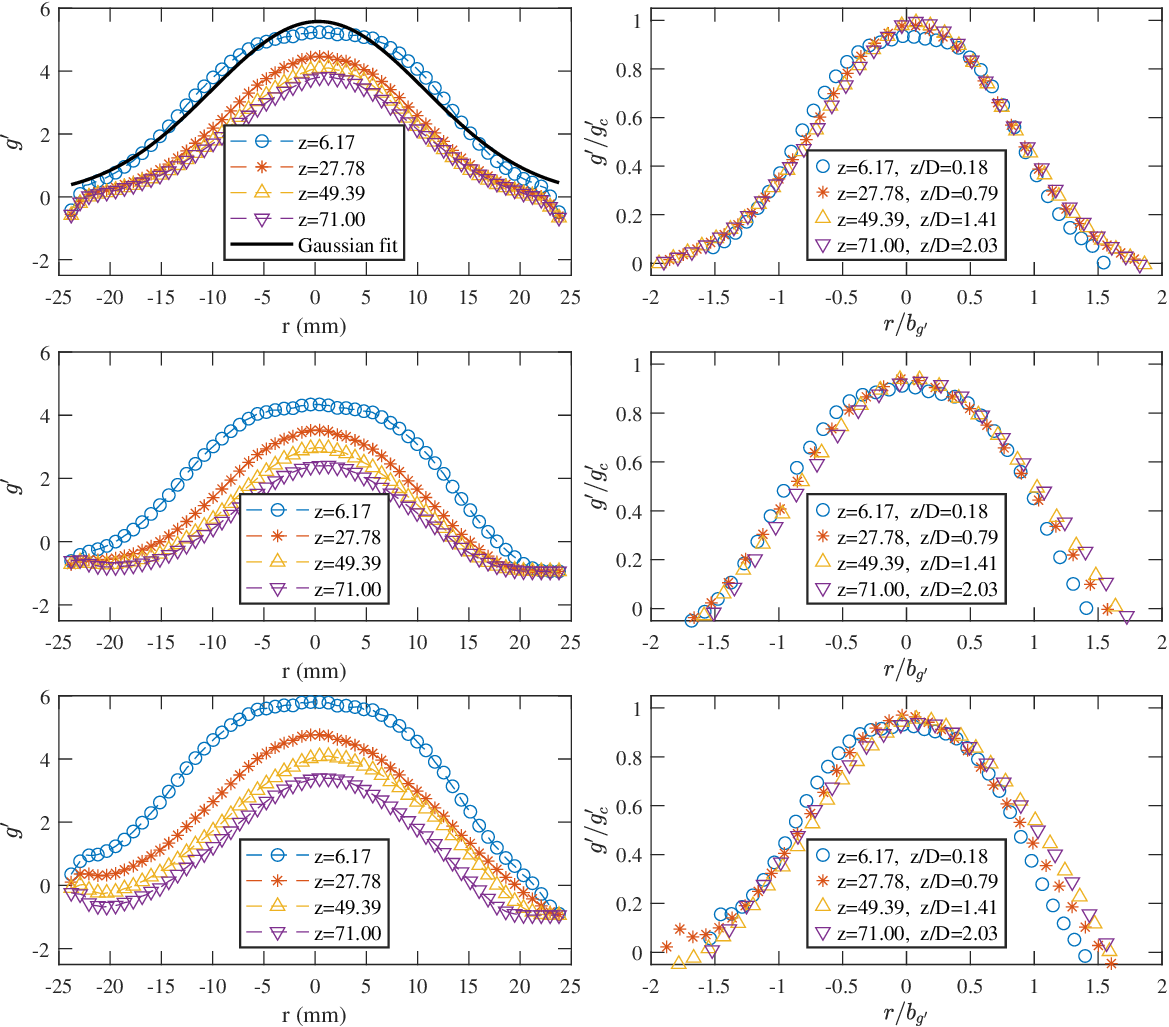}
\caption{Reduced gravity profiles extracted at different z planes (left column) and their collapse due to nondimensionalisation with $g^\prime _c$ and $b_{g^\prime}$ (right column) for Releases -I (top row), II (middle row), and III (bottom row). Solid black line shows the Gaussian curve fit to the $g'$ profile extracted at $z/D = 0.18$ for Release-I.}
\label{fig:gausScaling} 
\end{figure*}

To fit a Gaussian profile to the extracted centreline $g'$ data, MATLAB's `fit' function with `gauss1' model is used. A representative fit for the Release-I and profile extracted at $z/D = 0.18$ is shown in Fig.~\ref{fig:gausScaling}, where the coefficients of fitting are obtained as $g'_c = 5.59$ and $b_{g'}=14.86$ mm with a goodness of fit parameter \citep{MATLAB_Goodness_of_Fit}, $R^2 =0.98$, indicating a good quality of fit. The fitting parameters for the remaining profiles extracted are given in Table~\ref{tab:gausFitParams}. For the three plume releases, the profiles are extracted at four locations within the range z/D = 0.18 to z/D = 2.03. For each profile, $g'_c$ consistently decreases in the downstream direction, indicating an increase in density due to the entrainment of heavier ambient air. The values of $b_{g'}$, the radial scaling, also decrease in the downstream direction, showing the narrowing Gaussian spread due to a decrease in $g'$, which corroborates the increase in density from ambient air entrainment. 

From the fit results given in Fig.~\ref{fig:gausScaling}, it can be seen that the downstream profiles collapse tightly onto a Gaussian curve and hence are self-similar \citep{turner1979buoyancy, ezzamel2015dynamical,  parker2020comparison, milton2021entrainment, richardson2022entrainment, pang2025measurements}. However, the initial profile at $z/d=0.18$ exhibits a flat-topped shape characteristic of a potential core, rather than the peak of a Gaussian.

\subsubsection{Comparison with Similarity Solution and Simulation Data from the Literature}
We further compare the centreline reduced gravity ($g'$) and the characteristic plume radius (half-width), $b_{g'}$, obtained from TBOS reconstructions with the similarity solution and the simulation data available in the literature. The half-width, $b_{g'}$, is taken up to the point where the buoyancy drops to $1/e$ (where $\ln (e) = 1$) of the centreline value. The similarity solution provided in the literature by \cite{van2010universal, hunt2011classical} for  $g'$ and $b_{g'}$ is given as:     
\begin{align}
g' &= \left( \frac{4}{3 \pi^2} \right)^{1/3} \left( \frac{5}{6 \alpha} \right)^{4/3} B_0^{2/3} z^{-5/3}, \label{eq:similarity_gdash} \\
b_{g'} &= \frac{6 \alpha}{5} z, \label{eq:similarity_bg}
\end{align}

However, these solutions are derived for the plumes emanating from ideal point sources, and to apply them for the plumes from finite area sources, a source correction must be applied. Several methods for obtaining source correction have been reported in the literature. The method proposed by \cite{van2010universal} is used in the present work.  For lazy plumes ($ \Gamma_0 > 1 $), the virtual source location $ z_{\text{avs}} $ is given by:
\begin{align}
z_{\text{avs}} &= \frac{D}{8 \alpha} \zeta_{\text{avs}},  \\
\zeta_{\text{avs}} &= -\frac{10}{3} \Gamma_0^{-1/5} + \Gamma_0^{-1/5} \sum_{n=1}^{\infty} \prod_{i=1}^{n} \left[ \left( -\frac{4}{5} + i \right) \right. \notag \\
&\quad \left. \times \frac{1}{n - \frac{3}{10}} \frac{1}{n!} \left( \frac{\Gamma_0 - 1}{\Gamma_0} \right)^n \right], 
\label{eq:zeta_avs}
\end{align}
where $ \Gamma_0 $ and $ \alpha$ are the plume function and entrainment coefficient, respectively. While evaluating Eq.~\ref{eq:zeta_avs}, the summation is truncated when the next term to be added falls below $ 10^{-15} $. The $g'$ and $b_{g'}$ for the similarity solution are obtained using the source-corrected axial location and then plotted with the axial location starting from the physical source.  

The large-eddy simulation data from \cite{zhou2001study} were first transformed to align with the parametric expressions used in the present study, given in Table~\ref{tab:cases}. The density ratio corresponding to the two cases considered in his simulations is obtained from the reported temperature values. For the direct numerical simulation data from \cite{meehan2023richardson}, we first calculate the acceleration due to gravity, $g$, using the given $Ri$ and $Re$ as $Ri \times Re^2$, and then source velocity, $V_0$, using $Ri/Re$. These are the two parameters varied in their study, while keeping all other parameters fixed. Knowing the values of $g$ and $V_0$, the cases are characterized similarly to those in the present work.

\begin{figure*}[ht]
\centering
\includegraphics[width=0.95\textwidth]{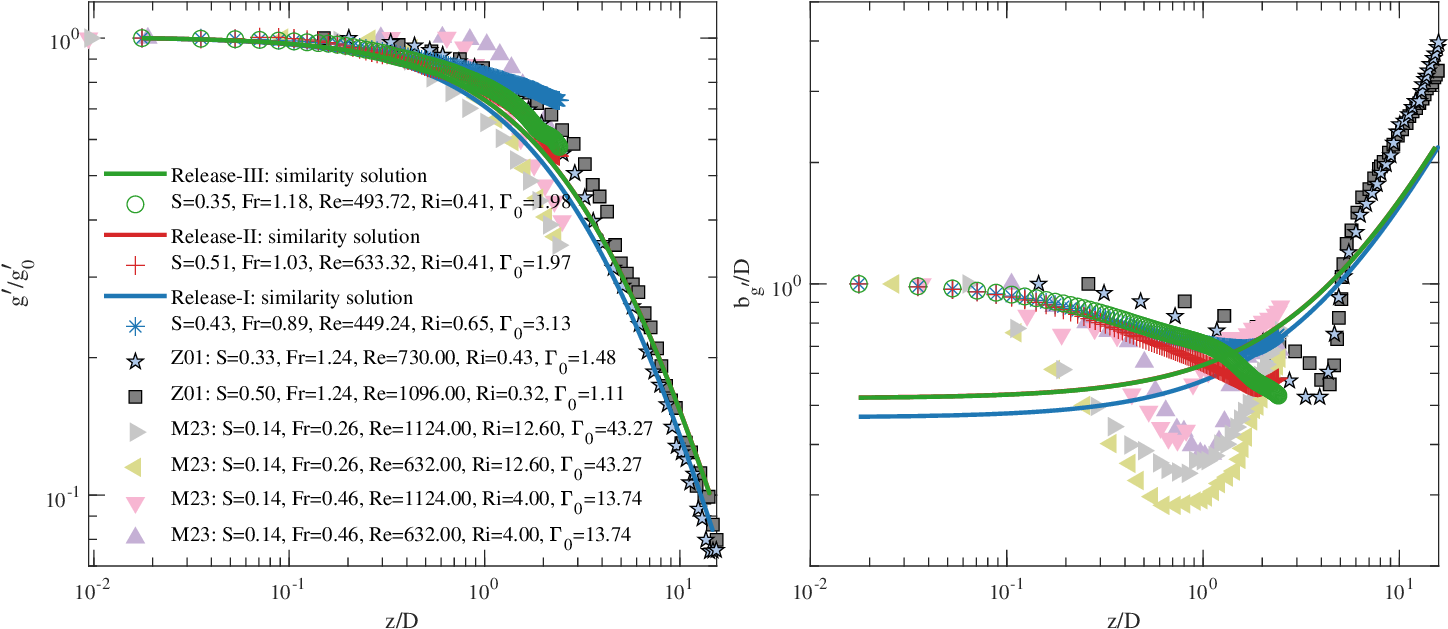}
\caption{Comparison of TBOS measurements with the corresponding similarity solution for the three release cases, along with the simulated data from literature. Solid lines denote similarity solutions, and Open marker symbols represent TBOS reconstructed data. Legend entries prefixed with 'Z01' and 'M23' correspond to data from \cite{zhou2001study} and \cite{meehan2023richardson}, respectively.}
\label{fig:similaritySoltnComp} 
\end{figure*}

The source parameters in the cases reported by \citet{zhou2001study} closely match the plume releases considered in the present experiment, resulting in good agreement as illustrated in Fig.~\ref{fig:similaritySoltnComp}. In contrast, the data from \citet{meehan2023richardson} correspond to significantly higher values of the source parameter ($\Gamma_0>13$) compared to the present study ($1.97<\Gamma_0<3.13$). Consequently, they exhibit significant quantitative deviation from the current measurements, although they demonstrate a similar qualitative trend- specifically, the pronounced necking characteristics of highly lazy buoyant plumes. These data points have nonetheless been included for reference and context.

The behaviour of Release-II (red) and Release-III (green) closely mimics the `Z01' case ($\Gamma_0$=1.48, filled star) as their source parameters ($\Gamma_0$ =1.97, 1.98) are comparable. Although $\Gamma_0$ is the dominant parameter governing the flow evolution, slight differences in other source properties prevent an exact overlap; nonetheless, both the $g'$ and $b_g$ trends follow the `Z01' dataset with reasonably good fidelity.

As shown in the plume width evolution, these `lazy plumes' exhibit non-self-similar behaviour in the near-field, approximately 4-6 source diameters from the origin \citep{hunt2005lazy, marjanovic2017evolution}. In this region, the excess buoyancy flux at the source drives a contraction, or `necking', of the plume profile. Downstream of this establishment zone, where the local Richardson number approaches unity, the flow transitions to a self-similar state described by the similarity solution. Since the current measurements are confined to the near-field establishment region only, a match with the similarity solution is not expected. Though a large deviation from the similarity solution is not expected either, since the plumes are weakly lazy. The data from \citet{zhou2001study}, which extend into the far-field, align well with the similarity solution computed using the current source conditions, validating the theoretical prediction for the asymptotic plume behaviour.

\subsection{Visualization of the Puffing Plume from the Reconstructed 3D Density Field}
Volumetric visualization of the instantaneous density fields can illustrate the plume dynamics and its axial, radial, and azimuthal variations, which differ substantially from the time-averaged density field presented in Fig.~\ref{fig:validatn}. This approach contrasts with the conventional 2D Planar Laser Induced Fluorescence (PLIF) and integral optical visualization methods such as schlieren and shadowgraphy. If the instantaneous volumetric density field from the current work could be combined with the velocity field, the plume dynamics and their entrainment characteristics could be accurately captured. Such detailed measurements can help develop better analytical models for entrainment. Additionally, these measurements can also provide the dynamics of the turbulent-non-turbulent interfaces of flow where buoyancy effects are present. In the present work, we illustrate the nature of the instantaneous density field for the three release cases described below.

\begin{figure*}[h]
\centering
\includegraphics[width=0.7\textwidth]{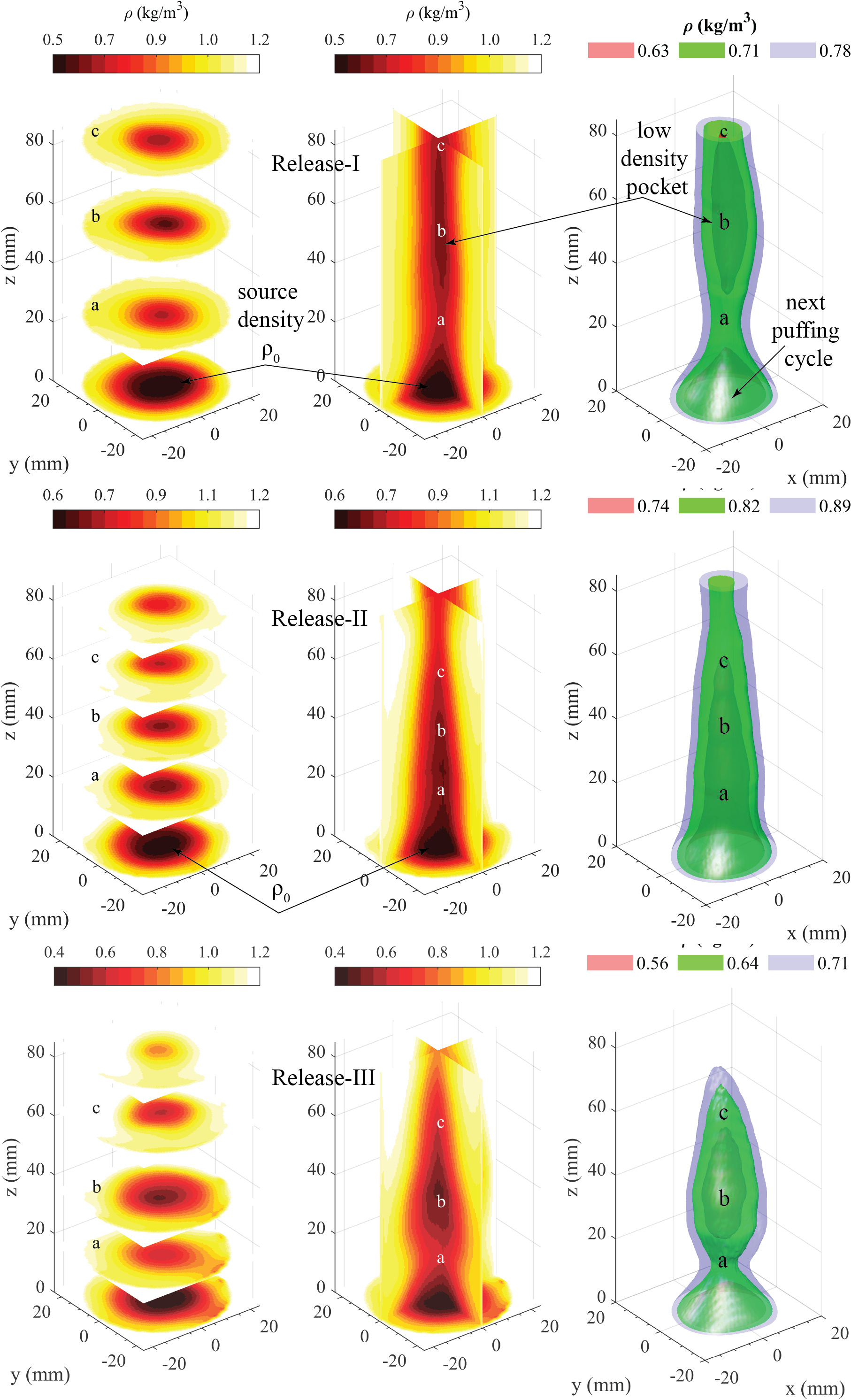}
\caption{Volumetric visualization of the 3D density field at selected time instances for three release cases at the measurement domain (scale: 0.0773/2 mm/px). The middle row shows the slightly forced plume (Release-II), where no Low Density Pocket (LDP) is observed. The top and bottom rows show Release-I and Release-III, respectively, highlighting the puffing phenomenon with detached LDPs. Each row includes density slices in transverse (left column) and longitudinal (middle column) planes and iso-surfaces (right column) of density.}
\label{fig:combInstDemo} 
\end{figure*}

Fig.~\ref{fig:combInstDemo} (middle row) shows a representative time instance for Release-II, arbitrarily selected, where the iso-surface (the right column) of the minimum density remains attached to the source. The two higher-density iso-surfaces surrounding the low-density iso-surface denote a gradual increase in density due to ambient fluid entrainment. Three points, marked as `a', `b', and `c' on the iso-surface visualization, are examined in the transverse (left column) and longitudinal (middle column) density planes. For Release-II (middle row) the longitudinal density planes reveal a gradual density increase within the innermost iso-surface (region around points `a' and `b'). The acceleration effect due to buoyancy that arises from the density difference is clearly visible from the iso-surface plot and the longitudinal variations. The transverse density slices at points `a', `b', and `c' show the azimuthal density variation, with a gradual density increase in both radial and downstream directions. Release-II does not exhibit a separated Low Density Pocket (LDP) of source fluid, unlike the remaining two releases as described below. This confirms the absence of puffing, as expected from Fig.~\ref{fig:neutralCurve} for Release-II, and a sustained, gradual entrainment of heavier ambient fluid into the plume in the downstream direction is observed.

For the lazy plumes in Release-I and Release-III, the puffing phenomenon characterized by the periodic formation and downstream convection of large-scale toroidal vortices \citep{cetegen1996experiments, Bharadwaj_Das_2017, bharadwaj2019puffing} is captured. As the toroidal vortex, forming near the source, moves downstream, it grows in size and pinches \citep{xia2018vortex} the plume fluid, inscribed by the toroidal vortex core, around the plume axis.
The Low-Density fluid Pockets (LDPs), generated by the plume, pinch off and travel downstream. The representative instances with the detached LDP are shown in Fig.~\ref{fig:combInstDemo}, top and bottom rows, for Release-I and Release-III, respectively. 

The LDP is annotated for Release-I, along with the three points- `a', `b', and `c' to examine the density in the transverse and longitudinal planes. The point `b' is taken within the marked LDP, while points a' and `c' are upstream and downstream of it, respectively. The density variation around the innermost iso-surface representing LDP can also be seen in the longitudinal slices (middle column). The density of the plume fluid increases towards the neck upto point `a' from the source and remains low only in a thin region around the plume axis at point `a', at this instant. The higher density at point `a' can also be confirmed by the transverse plane at `a' (left column). Moving from point `a' to point `b' within the marked LDP, a relatively low-density fluid is encountered at point `b'. The density increases radially outwards within the LDP as seen in the transverse slice `b'. Moving further downstream from point `b' to `c', the LDP shrinks and vanishes before point `c'. The variation of density along the plume axis from `b' to `c', as seen in the transverse slices, shows a rise in density. The transverse slices also confirm that the density field increases radially outward, as expected due to heavier ambient air entraining into the plume.  

The puffing phenomenon for Release-III, as shown in the bottom row of Fig.~\ref{fig:combInstDemo}, is expected to be more prominent than in Release-I. It can be seen that the maximum radius of LDP at point `b' and the minimum radius of neck at point `a' are significantly larger and smaller, respectively, in comparison to Release-I. However, for the ellipsoidal LDP region, the major diameter (aligned to the plume axis) has now shrunk with respect to the minor axis (aligned radially). The density within the ellipsoidal LDP region decreases towards the centre, as seen in both longitudinal and transverse density planes. In this release, compared to Release-I, the density increases due to ambient air entrainment in the downstream direction with a steeper rate, as seen in the longitudinal and transverse density slices. 

It is to be noted from the instability equations \citep{Bharadwaj_Das_2017} that puffing depends on $Re$, $Fr$, and $S$. The dependency of $Re$ is insignificant on puffing and its frequency. Thus, $Fr$ and $S$ become the most important parameters, and the puffing is governed by the density difference and gravity \citep{Bharadwaj_Das_2017}. Hence, the Release-III, which has a lower $Ri$ (a combination of $Fr$ and $S$) compared to the Release-I, shows a higher degree of puffing as observed from the density measurements. 

\clearpage

\section{Conclusion}\label{sec:conc}
This paper presents the first volumetric density measurements of buoyant plumes of a He–Air mixture using Tomographic Background-Oriented Schlieren (TBOS), with the primary objective of studying plume dynamics, particularly the visualization of the puffing phenomenon. A MATLAB-based code module, developed in-house, is employed for data processing and tomographic reconstruction computations. Three lazy buoyant plume configurations are examined as test cases. The transverse profiles of reduced density extracted at different heights above the source scale well when modelled as Gaussian, consistent with several experimental studies reported in the literature. Moreover, the centerline reduced gravity and plume half-width are compared with the similarity solution and simulation data available in the literature, showing good agreement. We further demonstrate that density measurements can be used to study plume dynamics using iso-surfaces. We also observe the motion of low-density pockets during puffing in the lazy plume releases, which is absent in the non-puffing plume. However, a lack of accompanying 3D velocity field information limits the utility of the 3D density field for quantitative statistics such as the entrainment coefficients. The 3D density field can be converted into a 3D velocity field using data-assimilation approaches. Moreover, synchronized 3D velocity measurements using tomographic PIV and Particle Tracking Velocimetry (PTV) could enable a complete analysis of such density-driven flows. The developed technique will be extended, along with 3D velocity measurements, in future studies to investigate the dynamics of buoyant plumes more comprehensively.

\bmhead{Acknowledgments}
The authors acknowledge the contributions of Kuchimanchi K. Bharadwaj and Karthik Murthy in developing the buoyant plume generation setup. The first author also gratefully acknowledges Professor Naren Naik (Department of Electrical Engineering and Centre for Lasers and Photonics (CELP), IIT Kanpur) for the course EE 659: Computational Aspects of Tomographic Imaging, which provided key conceptual insights relevant to this work.

\section*{Declarations}
\begin{itemize}
\item Funding: The corresponding author acknowledges the Bhabha Atomic Research Centre (BARC), Department of Atomic Energy, Government of India, for their partial support in the development of the TBOS system.
\item Conflict of interest: There are no competing interests to be disclosed
\item Ethics approval and consent to participate: Not applicable
\item Consent for publication
\item Data availability 
\item Materials availability
\item Code availability 
\item Author contribution: JM - Conducted experiments, developed the analysis software module, analysed data, and wrote the first draft; DD - Conceptualised the research, provided supervision and guidance, reviewed and revised the manuscript
\end{itemize}

\noindent


\begin{appendices}
\section{BOS Principle}\label{appB:bosFunda}

\begin{figure*}[h]
\centering
\includegraphics[width=0.9\textwidth]{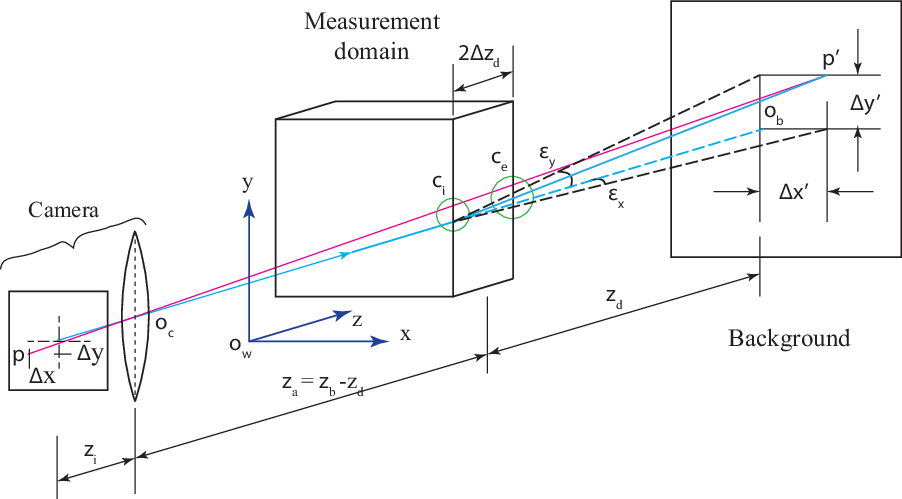}
\caption{Schematic of the optical setup illustrating the principle of background-oriented schlieren (BOS) measurements}
\label{fig:bosOptSetUp} 
\end{figure*} 

When a light ray passes through the measurement domain with a variable refractive index, $ n(x,y,z)= 1 + G \rho(x,y,z)$, it refracts according to the local value of $n$ as shown in Fig.~\ref{fig:bosOptSetUp}. The parameters $G$ and $\rho$ are the Gladstone-Dale constant and local density. A ray point on the background, $o_b$, in the absence of any density gradient, is getting imaged on the centre pixel of the image plane. Here, only the central ray from a pencil of light rays (light rays emitted by a point on the background in the conical shape) emanating from point $o_b$ is shown. 

With the variable density in the domain, the ray passing through it gets refracted, and to the camera, it seems to be coming from point p$^\prime$ on the background. This apparent point p$^\prime$ gets imaged as point p, on the image plane. The displacement from $o_b$ to p$^\prime$ can be decomposed into Cartesian components- $\Delta x' $, $\Delta y'$. These displacement components are caused by the density gradient in the x and y directions, $\partial \rho /\partial x$ and $\partial \rho /\partial y$, respectively. In the following paragraph, we develop the relation between the two using the $y$ component; a similar relation for the $x$-component will be inferred. 

It is assumed \citep{merzkirch2012flow, greivenkamp2004field} that the deviation of the ray from the z-axis (optical axis of camera) is negligibly small, therefore the ray remains straight within the measurement domain, but exits with a non-negligible curvature at the exit plane. In other words, the line inside the domain does not deviate from its straight-line path but exits with deflection angles $\epsilon _x$ and  $\epsilon _y$. The ray might encounter these negligible path deviations due to refractive index gradient (density gradients) anywhere along its path within the measurement domain. This assumption would be more accurate as the $2 \Delta z_d$ gets smaller and smaller as compared to the $z_b$, resulting in negligible ray path deviation with respect to its assumed straight path. Circles marked with $C_i$ and $C_e$ denote the entry and exit points of the representative ray, exaggerated here to emphasise that some path deviation is encountered, due to ray deflection, in real cases. Therefore, under the assumption, the circles will collapse to their respective centre points.

The assumption of paraxial optics \citep{venkatakrishnan2004density, van2014density} has also been taken, for the validity of which, $2\Delta z_d << z_d$. Due to this, a small curvature in the ray at the exit plane, $\epsilon _y$, leads to a detectable deflection on the imaging plane due to the long ray path. This is the so-called far-field BOS assumption; in cases where the background is not sufficiently far away (near-field BOS), a correction is required, as given by \cite{van2014density}. 

For the refraction of the ray, the simplified Euler-Lagrange equation results in: 

\begin{equation}
\epsilon_y = \int_{z_d - \Delta z_d} ^{z_d + \Delta z_d} \frac{1}{n_0}\frac{\partial n}{\partial y} \left( =G \frac{\partial \rho}{\partial y}  \right)dz,
\label{eq:euLagAppB}
\end{equation}
where $\partial n / \partial y = G \, \partial \rho/ \partial y $ is obtained from the Gladstone-Dale relation and $n_0$ is the refractive index of the undisturbed surrounding medium (ambient air in the current measurements).

Also, from the point p$^\prime$ on the background, assuming a small value of $\epsilon _y$:

\begin{equation}
tan(\epsilon _y)\approx \epsilon _y = \frac{\Delta y^\prime}{z_d} 
\label{eq:geomDef}
\end{equation}

Now using Eqs.~(\ref{eq:euLagAppB}) and (\ref{eq:geomDef}): 

\begin{equation}
\int_{z_d - \Delta z_d} ^{z_d - \Delta z_d} \frac{\partial \rho}{\partial y}dz = \frac{n_0 \Delta y'}{G \, z_d}
\label{eq:bosFinal}
\end{equation}
We use the above Eq.~(\ref{eq:bosFinal}) for our MATLAB-based software module, where $\Delta y^\prime$ are directly obtained by PIV computations using PIVlab. The calibration (scaling) factor is determined using a linear ruler placed on the background pattern.

Using the similarity triangle for the imaging geometry of p$^\prime$, the displacement in the vertical direction on the background $\Delta y^\prime$ and on the image plane $\Delta y$ can be related as:

\begin{equation}
\frac{\Delta y}{z_i}= \frac{\Delta y^\prime}{z_b}
\label{eq:yyDash_fov}
\end{equation}

Substituting $\Delta y^\prime$ in Eq.~(\ref{eq:bosFinal}) by $\Delta y$ from Eq.~(\ref{eq:yyDash_fov}) results in: 

\begin{equation}
\int_{z_d - \Delta z_d} ^{z_d + \Delta z_d} \frac{\partial \rho}{\partial y} dz= \frac{n_0}{G z_d} \left( \frac{z_b}{z_i} \Delta y    \right)
\label{eq:bosFinalCam}
\end{equation}

The Eq.~(\ref{eq:bosFinalCam}) \citep{merzkirch1987flow, raffel2000applicability, richard2001principle, meier2002computerized, venkatakrishnan2004density, raffel2015background}, explicitly includes the camera parameters.
The parameter $z_i$ can be replaced by the focal length $f_c$ using the thin lens formula $1/z_b + 1/z_i =1/f_c$; since $z_b >> z_i$, $1/z_i = 1/f_c$. However, care should be taken while using this simplification of $z_i$ being equal to $f_c$. While this approximation is valid in the limiting case, it should be used with caution.

Using Eq.~(\ref{eq:bosFinalCam}), the sensitivity, for a given density gradient ($\nabla \rho= \sqrt{(\partial \rho / \partial x)^2 + (\partial \rho / \partial y)^2}$) what would be the total deflection on the image plane ($\Delta_{r}= \sqrt{\Delta x^2 + \Delta y^2}$), can be given as:
 
\begin{equation}
\frac{\Delta_{r}}{\nabla \rho}=\left(\frac{G z_d}{n_0}\right) (2 \, \Delta z_d) \left(\frac{z_i}{z_b}\right) .
\label{eq:sensitivity}
\end{equation}

From Eqs.~(\ref{eq:bosFinal}) and (\ref{eq:yyDash_fov}), the minimum measurable density gradient is given as: 
\begin{equation}
\nabla \rho |_{min}=\left(\frac{n_0}{G z_d}\right) \left(\frac{1}{2 \, \Delta z_d}\right)\Delta^\prime_{r}|_{min} \left( = \frac{z_b}{z_i} \, \Delta_{r}|_{min}\right)
\label{eq:delRhoMin}
\end{equation}
 
where, $\Delta^\prime_{r}$ denotes the total deflection on the background, and $z_i = z_b f_c/(z_b - f_c)$. The noise floor in the cross-correlation-estimated deflection field was determined by processing 13 pairs of reference images from each camera, resulting in a mean RMS deflection of 0.08 pixels. Using the scaling factor at the background plane (0.0773 mm/px), this value corresponds to a physical deflection of 0.0064 mm. With $2\Delta z_d = 68.6$ mm, the sensitivity at the background plane is calculated as $\Delta^\prime_{r}/\nabla \rho = 9.40 \, [\mathrm{\mu m}/(\mathrm{kg}/\mathrm{m}^4)]$, which corresponds to a projected sensitivity at the sensor plane of $\Delta_{r}/\nabla \rho = 0.72 \, [\mathrm{\mu m}/(\mathrm{kg}/\mathrm{m}^4)]$.  The minimum detectable density gradient is calculated using Eq. (\ref{eq:delRhoMin}) as $\nabla \rho_{\min} = 0.68\,\mathrm{kg}/\mathrm{m}^4$.

\section{Camera Alignment Procedure and Coordinate System Validation} \label{append:camAlignErrors}
The cameras are rigidly mounted on heavy base stands. The nozzle is mounted at the centre of a rigid iron table, with a hole at the centre, within a vertical slot in the settling chamber. The nozzle can be adjusted in height. A paper printed protractor with angular markings at 0.5 degrees is pasted on the table, through which the nozzle protrudes from a cut at the centre. The protractor markings guide the crossline laser to precise angles. The camera, nozzle, and background are first positioned using floor construction lines based on FOV calculations, spirit levels (FREEMANS), and a cross-line self-levelling laser (STANLEY CL90). The background pattern plane, lens plane and sensor plane are parallel to each other. Hence, only a scaling factor is sufficient for calibration.

The vertical alignment of the nozzle is ensured by placing the spirit level at two perpendicular directions on the nozzle exit plane, thus constraining two rotational Degree Of Freedom (DOF). The circular geometry constrains the remaining rotational DOF, and the fixed mount constrains all translational DOFs. The centre of the nozzle, fixed in space, acts as the origin of the world coordinate system. 

Firstly, the camera is focused on the nozzle. The crossline laser mounted on the table issues a horizontal beam that intersects the camera mounting stand, providing a reference for adjusting the camera height relative to the nozzle exit plane. The beam passes through the nozzle centre and is aligned at the targeted camera angle (eg, 22.5 degrees) using the protractor. The translational DOF associated with the camera's height relative to the nozzle exit plane is constrained by fixing the camera's height. The camera is adjusted so that the nozzle centre, computed as the mean of the nozzle edge positions, lies at the centre of the camera frame, restraining the yaw rotational DOF and the sideways translational DOF. The camera roll and pitch are arrested by placing the spirit level at two perpendicular positions on the camera body; the nozzle image height and the alignment of the nozzle image edges with the camera frame further confirm this. The width of the nozzle image on the camera frame fixes the radial distance between the nozzle and camera, and constrains the remaining 3rd translational DOF. If the nozzle and camera are closer than the design distance, the nozzle image spans more horizontal pixels than expected from the FOV calculations. Cameras are adjusted until the centred nozzle image spans the same pixel width in all views.

For the final acquisition, the camera is gently focused on the background and aligned as follows. The parallelism between the camera lens and the background is achieved by drawing a rectangle on the background and adjusting its orientation until its sides align with the camera frame edges and fill the maximum image height and width as per the FOV calculations. Afterwards, a steel ruler was gently placed against the background pattern, and images were acquired in different orientations. These images were used to calculate the scaling factor. If the orientation and the distance of the background pattern from the camera are correct, the scaling factor obtained from different views should be consistent. The steel ruler images were also used to check for radial lens distortion by fitting straight lines to the ruler edges and evaluating the residuals.

\begin{figure}[h]
\centering
\includegraphics[width=0.5\textwidth]{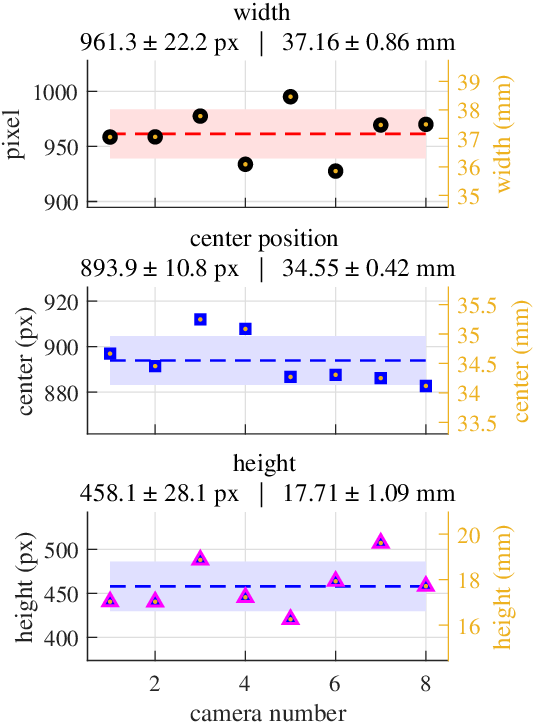}
\caption{Quantification of coordinate system uncertainty. Dashed lines show the mean across cameras, the shaded band represents the standard deviation, and markers indicate the values for individual cameras. The numerical values of mean and standard deviation are indicated in the title of each panel.}
\label{fig:camCalibrationErrors} 
\end{figure}

Any misalignment due to retouching the lens for re-focusing on the background is quantified by processing the 12 reference images with the blurred nozzle from each camera. Pixel intensity profiles across the blur transition are extracted. By fitting the Edge Spread Function (ESF) to the transition from non-blur to blur region, the width, height and centre of the blur nozzle image were computed. The mean and standard deviation of the width, centre and height of each camera are plotted in Fig.~\ref{fig:camCalibrationErrors}. The standard deviation from the mean for these quantities is 22.2 px (0.86 mm), 10.8 px (0.42 mm), and 28.1 px (1.09 mm), representing the uncertainty in the camera alignment. Finally, prior to the experimental run, the nozzle was physically lowered to achieve a downward shift of approximately 200 pixels in the image frame, thereby maximising the available field of view for capturing the plume flow field.

This camera positioning method is conceptually simple and, based on the authors’ experience, naturally aligns with the parallel-beam tomographic forward modelling used in this work, where only the angular spacing and a scaling factor for each camera are required. This is made possible by leveraging the specific geometric arrangement of the BOS rig, in which the cameras and backgrounds are mounted circumferentially around the measurement domain. However, the authors acknowledge that though the procedure is rigorous, it is manual and requires multiple iterations of physically adjusting the orientation of the cameras and backgrounds. Therefore, automated multi-camera calibration methods \citep{hartley2003multiple, nicolas2016direct, grauer2023volumetric, koponen2023background, barta2025comparison} that use a calibration target would benefit future implementations.

\section{Uncertainty quantification and assessment of accuracy}\label{appC:errorEstimation}
\subsection{Selection of background pattern}
A random-dotted pattern with a mean dot diameter of 5.7 pixels and a mean dot density of 0.0053 dots per-pixel-square was used. This results in 5.5 dots per interrogation window of 32x32 pixels (2.47 mm x 2.47 mm, at the pattern location), the smallest interrogation window of a three-pass approach used for the dot displacement calculations. The number of dots per interrogation window in the current measurement is in alignment with the recommendation of 5 for PIV \citep{raffel2018particle} and 7-10 by \citep{vinnichenko2012accuracy}, or 4-5 by \citep{gojani2013measurement} for BOS applications, as summarized by \cite{schwarz2023practical}, who suggest typically 6 dots per interrogation window.

\subsection{Estimation of blur circle or the centre of confusion (CoC)}
The blur circle was estimated using the expression provided by \citet{Greenleaf1950, lang2017measurement}, as shown in Eq.~(\ref{eq:CoC}), yielding a value of 0.29 mm on the sensor plane. The BOS experimental parameters used in the calculation are: $z_a = 600$ mm, $z_d = 600$ mm, $f_c = 85$ mm, and the f-number, $f_{\#} = 22$. 

\begin{equation}
CoC=\frac{f_c^2 z_d}{f_{\#}z_a (z_a + z_d -f_c)}
\label{eq:CoC}
\end{equation}

Taking the pixel size of 5.5 $\mu m$, it results in $\sim$53 pixels at the sensor plane. Using the scaling factor of 0.0773/2 (mm/px), it transfers to 2 mm at the phase object plane. Since the CoC is larger than the smallest interrogation window chosen, 32 px (1.23 mm) $\times $ 32 px for the dot displacement calculation, the spatial resolution of the current measurements would be mainly limited by the former. \cite{nicolas20173d} observed a similar limitation, reporting a CoC diameter of 2.5 mm compared to an interrogation-window resolution of 0.54 mm.

\subsection{Error due to exposure time }
In the lazy plumes, the velocity is expected to increase to approximately 2.85 times the source velocity \citep{zhou2001study, meehan2023richardson} at the neck region, reaching a maximum. The maximum source velocity is 0.69 m/s in release-III, which can be used to estimate the fluid element displacement during exposure. Therefore, the maximum estimated velocity of 1.97 m/s over an exposure of 100 $\mu$s leads to a displacement of 0.197 mm. Using the scaling factor at the central plain of the measurement domain (0.0773/2 mm/pixel), this results in a pixel displacement of 5 pixels. However, this displacement doesn't affect the accuracy of the presented density field, which is mainly dominated by the smallest Interrogation Window (IW) of 32$\times$32 pixels. Also, the relative displacement of the convecting density structures in the plume flow would likely remain the same within the IW, though encountering minor streak blur in the pattern images. Therefore, during the cross-correlation computation, the relative placement of the vectors representing the deflection of different rays is likely to remain consistent.

\subsection{Error due to lens distortions}
The use of aberration-free, aspherical, medium telephoto lenses, such as the 85 mm lens used in the current experiment, is considered to have negligible radial distortion. Consequently, lens distortion is often not explicitly corrected in studies using long-focal-length lenses, as reported in several planar PIV experiments \citep{raffel2018particle}. Nevertheless, we verified the radial distortion by fitting straight lines to the edges of the steel ruler images used to obtain the scaling factor from the background patterns. We found that the straight lines remain straight, signifying no radial distortions. Moreover, maximum pixel deviations were only of the order of 3-4 pixels near the frame edges. Since, in the present study, the sensor, lens, and image planes are parallel to each other, resulting in a planar PIV configuration, the lens distortion corrections were deemed unnecessary. Tangential distortion in modern scientific cameras is considered negligible \citep{devernay2001straight, tsai2003versatile, nowakowski2007lens} and was therefore not accounted for.

\subsection{Camera FOV and the reconstruction square size}\label{appC:camFOV}
\begin{figure*}[h]
\centering
\includegraphics[width=0.85\textwidth]{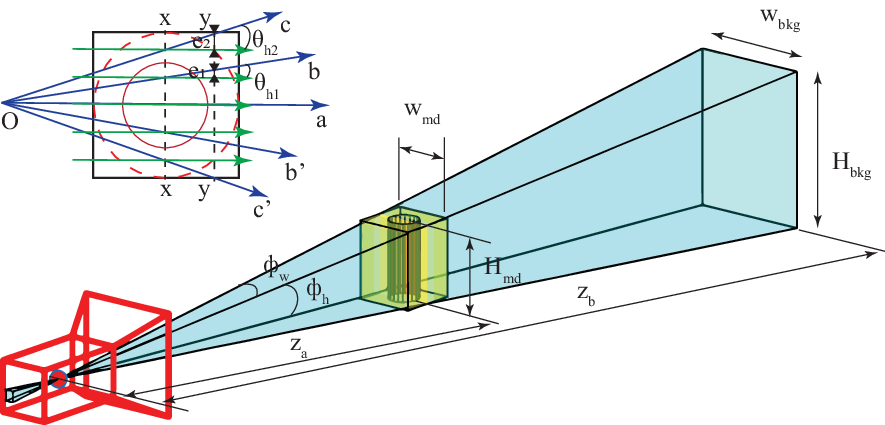}
\caption{Schematic showing the camera's field of view at the background and the central plane of the measurement domain. The inset illustrates the ray position error introduced by using parallel-beam forward modelling instead of fan-beam modelling, which accounts for the diverging rays from the camera. The divergence angles are exaggerated for clarity.}
\label{fig:camFOV} 
\end{figure*}

The cross-correlation of the background images provides the results on a spatial grid at an axial distance of $z_b$ from the optical centre of the camera using a scaling factor of 0.0773 mm/pixel. This spatial grid is scaled down by a factor of 2 to match the central plane passing through the cylindrical phased object at the centre of the targeted measurement volume. The Field Of View (FOV) (Fig.~\ref{fig:camFOV}) at the background location along the width and height of the camera frame are $W_{bkg}$ = 137.3 mm and $H_{bkg}$ = 182.6 mm, respectively. Using the $z_a$ = 600 mm and $z_b$=1200 mm, this scales down to a FOV of $W_{md}$ = 68.6 mm and $H_{md}$ = 91.3 mm at the central plane of measurement domain. The camera resolution of 1768 x 2352 with a pixel size of 5.5 micrometers results in a sensor size of (sensor width x sensor height) 9.7240 mm x 12.9360 mm. Therefore, the angular field of view, using the focal length $f_c$ = 85 mm, as $\phi/2 = \tan^{-1}  [\textrm{ sensor-width/2, \, sensor-height/2}]\,f_c$ = $[3.28, 4.36]^\circ$. Note that the angular fields are different along height and width due to the rectangular size of the sensor. Geometrically, the angular field of view represents a truncated cone, with a single value representing the angular FOV. Further, the FOV at the locations $z_a$ and $z_b$ along the optical axis from the camera optical centre can be calculated as $2 [z_a, z_b] \tan( \phi /2)$, respectively. 

The reconstruction square was defined as the largest square that could be inscribed within the cylindrical viewing domain spanned by the width of the FOV, denoted by diameter $ W_{md} $. This resulted in a square side length of $ W_{md}/\sqrt{2} = 48.51 \, \text{mm} $. A reconstruction square that circumscribes this cylindrical domain would include corner regions through which no rays pass, leading to poor reconstruction quality. Therefore, the inscribed square domain was chosen to ensure complete ray coverage within the reconstruction region.

\subsection{Error due to parallel beam forward modelling} \label{appSec:FovAndRayDivErr}
\begin{figure*}[h]
\centering
\includegraphics[width=0.9\textwidth]{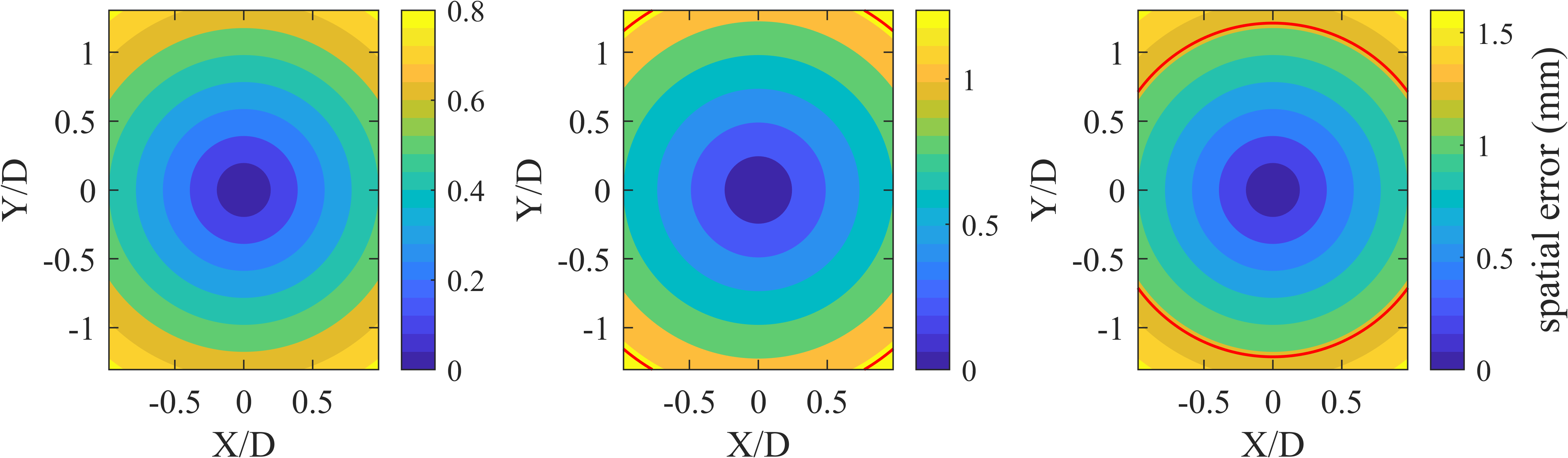}
\caption{The vertical planes showing the spatial error due to the assumption of parallel rays instead of diverging rays within the measurement domain, measured from the axis of the cylindrical phased object region, at the normalised radial positions (from left to right) of r/D = 0.25, 0.4, and 0.5.}
\label{fig:rayDivergErr} 
\end{figure*}

The spatial error due to ray divergence can be calculated by comparing the spatial position of a parallel ray to its actual position, diverging ray position. The angular field of view in the horizontal and vertical planes is given by $\phi_h$ and $\phi_v$, respectively. A schematic illustrating the error due to parallel beam modelling in the horizontal plane is given in the inset of Fig.~\ref{fig:camFOV}. The parallel rays (shown in green) are cast by hinging them to the central plane, xx, of the measurement domain as seen from each camera view. The difference in the path of these idealised parallel rays and the actual diverging rays increases moving away from the central plane. Therefore, the error due to the parallel beam model also increases and is equal to the offset distance from the central plane, multiplied by the tangent of the divergence angle of the ray. The central ray in the camera frustum coincides with the central parallel ray of the idealised forward projection model. The rest of the rays diverge outwards, and their angle of divergence increases moving radially outward from the central ray. These divergence angles for the two representative rays are indicated by $\theta_{h1}$ and $\theta_{h2}$. The spatial error due to this divergence is denoted by $e_1$ and $e_2$, respectively (Fig.~\ref{fig:camFOV} inset). A similar argument can be developed for the divergence in the vertical plane. The errors calculated in the horizontal plane and vertical planes by such ray divergences are added to calculate the total spatial error due to parallel beam projection. The calculated total spatial errors for three vertical planes taken at increasing radial distance from the central axis at $r/D$ values of 0.25,0.4, 0.5 of the phase object, represented by the cylinder inside the measurement domain (Fig.~\ref{fig:camFOV}), are shown in Fig.~\ref{fig:rayDivergErr}. The contour corresponding to the error of 1.23 mm, which is the minimum spatial resolution constrained by the interrogation window of 32 px $\times$ 32 px (1.23 mm $\times$ 1.23 mm) with 50\% overlap, resulting in a vector spacing of 16 pixels (0.61 mm) is shown by the solid red line. As can be seen, the error due to parallel beam modelling only increases after the radial distance of 0.5D, and hence does not dominate in the region where the phase object is present.

\section{Assessment of quasi-shadowgraph effects} \label{app:quasiShadowEffects}
\begin{figure}[h]
\centering
\includegraphics[width=0.5\textwidth]{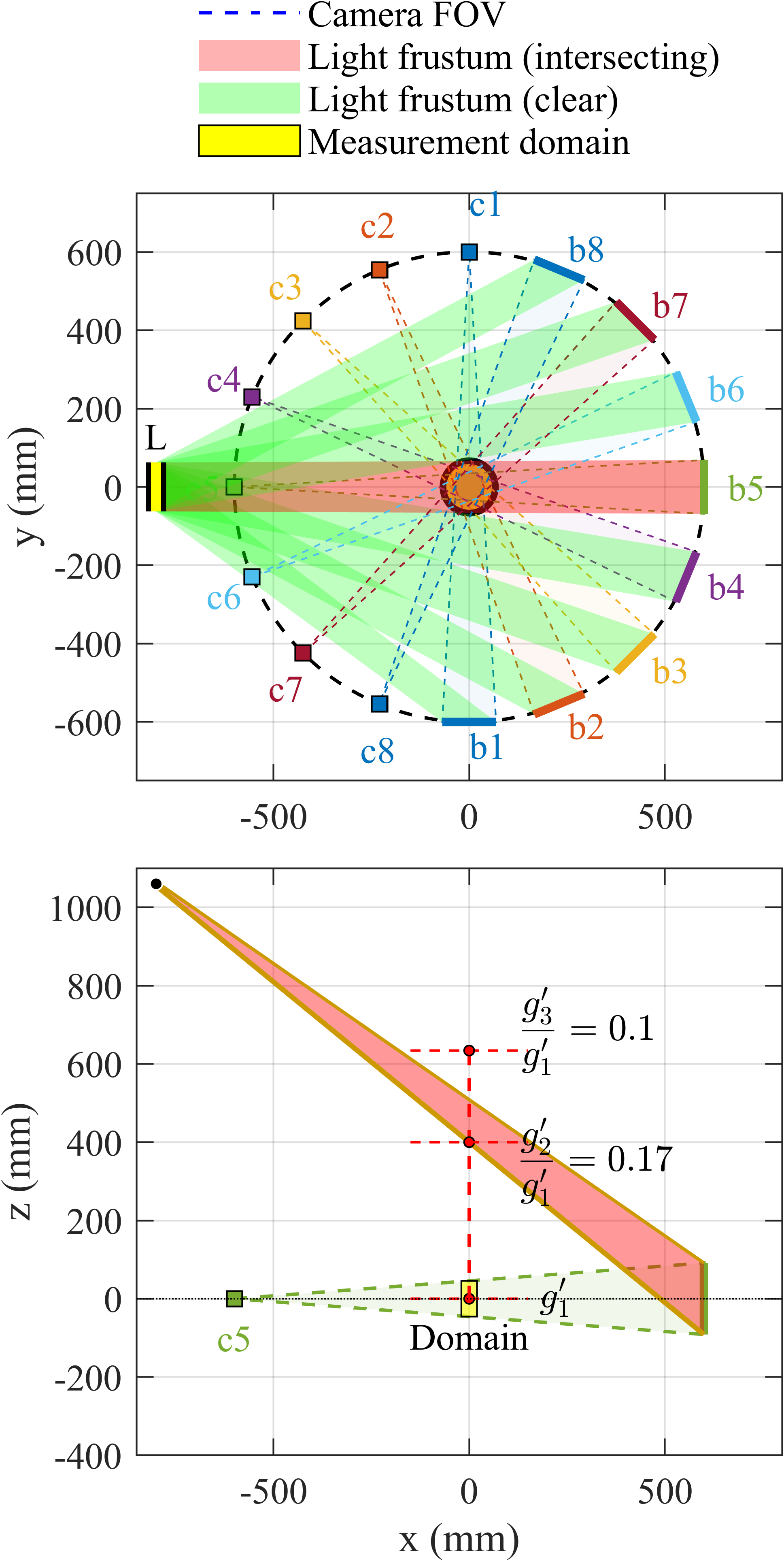}
\caption{Geometry of light frustum from source to each of the background patterns in the experimental rig. The symbols `L',`c', and `b' denote the light source, camera, and backgrounds, respectively. The top panel shows the top view, and the bottom panel shows the side view.}
\label{fig:quasiShadow} 
\end{figure}

The quasi-shadowgraph effects are assessed based on the position of the light source. The light source is a 1000-watt halogen tube with a 10 mm diameter and 125 mm length. It was mounted at an offset distance of 200 mm and a height of 1060 mm above the camera optical axis relative to camera 5. Figure~\ref{fig:quasiShadow} below plots the light frustum from the source to each background. The frustum is created by extending rays from the projected light-source rectangle corners to the corners of the background pattern. The top panel shows the top view; the bottom panel shows the side view. 

We estimate the relative magnitude of the reduced gravity, $\frac{g^\prime}{g^\prime_1}$, using the similarity solutions in Eqs.~(\ref{eq:similarity_gdash}) and (\ref{eq:similarity_bg}). For the radius of the measurement domain, shown in the top view, we use the plume radius at the z location where $\frac{g_2^\prime}{g_1^\prime}$ = 0.17. The top view shows that the light frustum mainly affects camera 5. In the side view, the frustum intersects the plume only at heights $z \ge z_2$, where the reduced gravity ratio is at most 0.17. This ratio decreases further to less than 0.1 at $z = z_3$. As the light frustum passes through the outer periphery of the plume, characterized by weak density gradients ($\le$17\% of the value at the FOV center at $z = z_1$) and lacking sharp discontinuities like shock waves, the resulting shadows on the background are negligible. 

The light source is distributed (line, not a point) and non-collimated, producing softer, more gradual shadows. Furthermore, any low-intensity shadows caused by weak density gradients would be removed during the PIVlab toolbox's image preprocessing steps- CLAHE and a high-pass filter. Consequently, the cross-correlation procedure will be dominated solely by the dot intensity distribution. 

\section{Details of ray deflection computation using Cross-Correlation approach}\label{appB:crossCorr}
A Fourier-transform-based approach for the cross-correlation computation, as implemented in PIVlab \citep{thielicke2014pivlab, raffel2018particle}, is used for ray deflection estimation. For the discrete images functions $g(p,q)$ and $h(p,q)$, representing the background images with and without the density gradients, respectively, and of resolution $P \times Q$ pixels, the discrete cross-correlation function $c(p', q')$ is calculated using the correlation theorem. The index ranges are \( p\in [0, P-1] \), \( q \in [0, Q-1] \), and the correlation is computed for shifts \( p' \in [-(P-1), P-1] \), \( q' \in [-(Q-1), Q-1] \).
According to the correlation theorem, the Fourier transform of the cross-correlation is given by
\begin{equation}
F[c(p', q')] = F[g(p,q)]F[h(p,q)],
\label{eq:corrTheorem}
\end{equation}

where $F[\cdot]$ denotes the two-dimensional Discrete Fourier Transform (DFT) of a discrete function. For example, the DFT of the function \( g(p, q) \) over a grid of size \( P \times Q \) is given by:

\begin{align}
&G(k_x, k_y) = \nonumber \\
&\sum_{p=0}^{P-1} \sum_{q=0}^{Q-1} g(p, q)  \exp \left\{ -2\pi i \left( \frac{k_x p}{P} + \frac{k_y q}{Q} \right) \right\}
\label{eq:DFT}
\end{align}

where \( k_x \in \{0, 1, \dots, P-1\} \) and \( k_y \in \{0, 1, \dots, Q-1\} \) are the frequency components in the \( x \)- and \( y \)-directions, respectively.

Therefore, the correlation function can be obtained by the inverse Fourier transform as 
\begin{equation}
c(p,q)= F^{-1}[F[g(p,q)]F[h(p,q)]],
\label{eq:fftCrosCor}
\end{equation}

where $F^{-1}[\cdot]$, the two-dimensional Inverse Discrete Fourier Transform (IDFT) used to reconstruct the original function $ g(p, q) $ from its Fourier transform $ G(k_x, k_y) $, is given by:

\begin{align}
&g(p, q) = \nonumber \\
&\frac{1}{P Q} \sum_{k_x=0}^{P-1} \sum_{k_y=0}^{Q-1} G(k_x, k_y)  \exp\left\{2\pi i \left( \frac{k_x p}{P} + \frac{k_y q}{Q} \right)\right\}
\label{eq:IDFT}
\end{align}
where \( g(p, q) \) is reconstructed at each point \( (p, q) \) in the spatial domain. In PIVlab, Eq.~(\ref{eq:fftCrosCor}) is used to compute the cross-correlation by taking the DFT of the two images, multiplying them, and then applying the IDFT to the product. 

The background pattern images with and without density gradients have been processed using the open source MATLAB toolbox PIVlab \citep{thielicke2014pivlab, thielicke2021particle} using the FFT based multi pass cross-correlation (Eq. (\ref{eq:fftCrosCor})) as described above to obtain the cross correlation based displacements. In PIVlab, image preprocessing was performed to enhance the quality of BOS images. Contrast-limited adaptive histogram equalization (CLAHE) was applied using a 64-pixel window to improve local contrast. A two-dimensional Wiener adaptive denoising filter with a 3-pixel window was employed to reduce noise. Additionally, an auto contrast stretch was implemented, which remaps the minimum and maximum intensity values of the input image to the full dynamic range (e.g., 0 to 255), while saturating the lowest and highest 1\% of input intensities to the extremes of the output range. Displacement fields were calculated using a fast Fourier transform-based cross-correlation algorithm with window deformation. The analysis involved three passes with interrogation areas of 128×128, 64×64, and 32×32 pixels, respectively, each with 50\% overlap. A Gaussian fit was used to estimate the sub-pixel location of the cross-correlation peak. The resulting vector grid had a spacing of 16 pixels in both directions. The details of the processing are also summarized in Table~\ref{tab:pivCalc}.

\begin{table}[h]
\centering
\caption{The details of the cross-correlation computation to obtain the ray deflection using the background pattern images with and without the presence of density gradients}
\label{tab:pivLab}
\begin{tabular}{@{}ll@{}}
\toprule
Image Preprocessing & CLAHE  64 px\\
 &Highpass filter 15 px\\
  &weiner denoising 3px\\
  & auto contrast stretch\\
Analysis method& FFT window deformation\\
Inter. area (pass 1)&128 $\times$ 128 pixels (50\% overlap)\\
Inter. area (pass 2)&64 $\times$ 64 pixels (50\% overlap)\\
Inter. area (pass 2)&32 $\times$ 32 pixels (50\% overlap)\\
Spurious vector filtering & velocity based validation\\
calibration & 0.0773 mm/pixel\\
\botrule
\end{tabular}
\label{tab:pivCalc}
\end{table}

The computed displacement vectors were further post-processed in the so-called vector validation step of PIVlab using a standard deviation filter threshold with a value of 8 that filters the vectors that are outside of the mean $\pm$ 8 $\times$ standard deviation and a local median filter with a threshold of 3, which applies a median filtering using the values in the 3x3 vector  neighbourhood of central vector. The Release-III, which has the lowest density ratio among the three releases, led to stronger astigmatism \citep{nicolas20173d, rajendran2020uncertainty, xu2025fringe} effects in regions with sharp density gradients, resulting in poor image quality and, consequently, spurious vector estimates. These artefacts were more pronounced along the plume axis. A spatial smoothing filter in PIVlab using the `smoothn' algorithm with robust argument \citep{garcia2010robust, garcia2011fast} was applied across all three cases to remove such outliers.

%
%
%

\section{Effect of sinogram interpolation and post-processing steps} \label{append:D}

Sparse-angle tomographic reconstructions often exhibit streak artefacts resulting from limited angular coverage. Sinogram interpolation is commonly employed to address these artefacts by estimating intermediate projections, as previously discussed in Sect.~\ref{sub:sinInt}.

\begin{figure*}[h]
\centering
\includegraphics[width=0.75\textwidth]{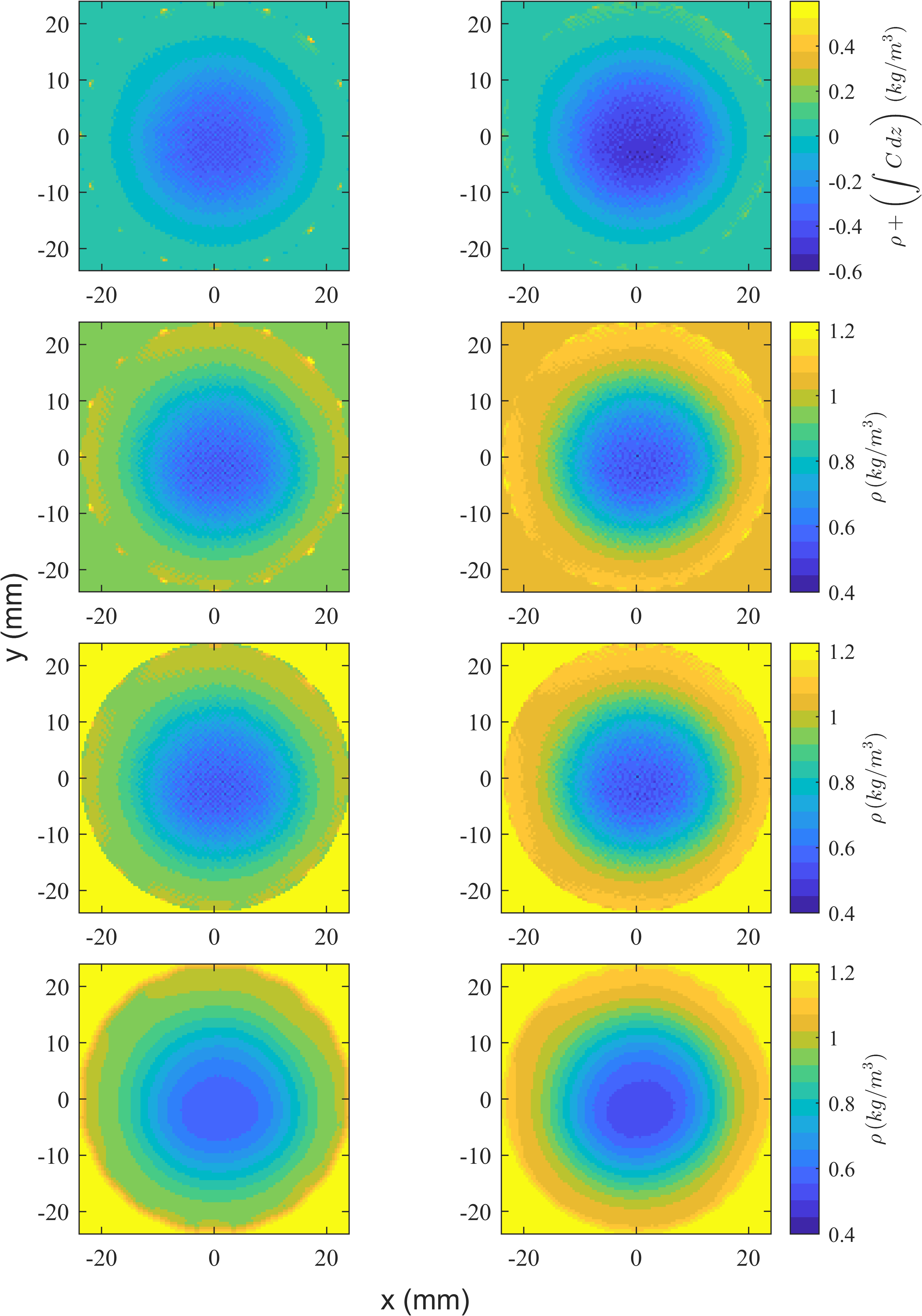}
\caption{Comparison of reconstruction results with and without sinogram interpolation for the $z$-plane at 6.17 mm (Plane-10) of Release-I, Frame 3. The left column shows data without interpolation, while the right column includes interpolation. Each row illustrates the effect of successive post-processing steps: (a) as obtained from SART reconstruction; (b) after applying constant correction using the source density at the nozzle exit centre; (c) after spatial filtering to reduce noise; (d) after applying a $5 \times 5 \times 5$ moving average filter to further suppress high-frequency noise.}
\label{fig:sngInterpAndPostProcessing} 
\end{figure*}

The impact of sinogram interpolation is demonstrated by comparing reconstructions with and without interpolation, using data from Release-I, Frame 3, at $z=6.17$ mm (Plane-10), as shown in Fig.~\ref{fig:sngInterpAndPostProcessing}. The left column presents reconstructions obtained directly from sparse-angle projections, while the right column displays results following sinogram interpolation. Each row corresponds to a distinct post-processing stage: (a) the raw reconstructed field, (b) application of undetermined constant correction resulting from Poisson integration with Neumann boundary conditions, (c) removal of non-physical values outside the reconstruction circle and above ambient density at the periphery, and (d) smoothing with a $5\times5\times5$ moving average filter to reduce high-frequency random noise.

Without interpolation, prominent dot-like artefacts are observed near the reconstruction circle periphery (rows a and b), whereas these artefacts are absent when interpolation is applied. Although subsequent filtering (row c) diminishes these artefacts and yields similar final fields in both cases, sinogram interpolation produces a physically consistent reconstruction prior to the constant correction step (row a). Based on the observed qualitative improvements in rows a and b, sinogram interpolation was implemented in this study. Nevertheless, a comprehensive quantitative assessment of its advantages remains an important area for future research.

\end{appendices}

\clearpage
\bibliography{sn-bibliography}



\end{document}